\pdfoutput=1
\documentclass{article}
\usepackage{algorithmic}
\usepackage{algorithm}
\usepackage{color}
\usepackage{graphicx}
\usepackage{subfigure}
\usepackage{amssymb, amsmath, amsthm}
\usepackage{graphics}
\usepackage{fullpage}

\newcommand{\IR}{IR}
\newcommand{\FULLIR}{Influence Rank}
\newcommand{\IE}{IE}
\newcommand{\FULLIE}{Influence Estimation}
\newcommand{\BP}{IP}
\newcommand{\FULLBP}{Influence Propagation}
\newcommand{\IRIE}{IRIE}
\newcommand{\FULLIRIE}{Influence Rank Influence Estimation}
\graphicspath{{figures/}}
\usepackage{authblk}

\date{}

\title{IRIE: Scalable and Robust Influence Maximization \\in Social Networks}

\author[1]{Kyomin Jung\thanks{kyomin@kaist.edu}}
\author[1]{Wooram Heo\thanks{modesty83@kaist.ac.kr}}
\author[2]{Wei Chen\thanks{weic@microsoft.com}}
\affil[1]{Korea Advanced Institute of Science and Technology}
\affil[2]{Microsoft Research Asia}

\begin{document}

\maketitle
\begin{abstract}
Influence maximization is the problem of selecting top $k$ seed nodes
	in a social network to maximize their influence coverage under
	certain influence diffusion models.
In this paper, we propose a novel algorithm IRIE that integrates a new message passing based
	influence ranking (IR), and influence estimation (IE)
	methods for influence maximization in both the independent cascade (IC)
	model and its extension IC-N that incorporates negative opinion propagations.
Through extensive experiments, we demonstrate that IRIE matches the influence
	coverage of other algorithms while scales much better than all other algorithms.
Moreover IRIE is more robust and stable than other algorithms both in running time and memory usage for various density of networks and cascade size.
It runs up to two orders of magnitude faster than other state-of-the-art algorithms such as PMIA for large networks with tens of
	millions of nodes and edges, while using only a fraction of memory
	comparing with PMIA.
\end{abstract}




\section{Introduction}

Word-of-mouth or viral marketing has long been acknowledged as an effective
	marketing strategy.
The increasing popularity of online social networks such as Facebook
	and Twitter provides opportunities for conducting large-scale online
	viral marketing in these social networks.
Two key technology components that would enable such large-scale online
	viral marketing is modeling influence diffusion and influence maximization.
In this paper, we focus on the second component, which is the problem
	of finding a small set of $k$ {\em seed nodes} in a social network to
	maximize their {\em influence spread} ---
	the expected total number of activated nodes after the seed nodes
	are activated, under certain influence diffusion models.

In particular, we study influence maximization under the classic
	independent cascade (IC) model~\cite{Kempe:MSI} and its extension
	IC-N model incorporating negative opinions~\cite{Chen:IC-N}.
IC model is one of the most common information diffusion model
	which is widely used in economics, epidemiology, sociology, and so
	on~\cite{Kempe:MSI}.
	Most of existing researches for the influence maximization problem are
	based on the IC model,
	assuming dynamics of information diffusion among individuals are independent.
Kempe et al. originally proposed the IC model and a greedy approximation
	algorithm to solve the influence maximization problem under the
	IC model~\cite{Kempe:MSI}.
The greedy algorithm proceeds in rounds, and in each round one node with
	the largest marginal contribution to influence spread is added
	to the seed set.
However, computing influence spread given a seed set is shown to
	be \#P-hard~\cite{Chen:PMIA}, and thus the greedy algorithm has to
	use Monte-Carlo simulations
	with a large number of simulation runs to obtain an accurate estimate of
	influence spread, making it very slow and not scalable.
A number of follow-up works tackle the problem by designing
	more efficient and scalable optimizations and
	heuristics~\cite{Kimura:TID,Leskovec:CELF, Goyal:CELF++, Chen:EIM,Chen:PMIA,Goyal:CELF++,JSCWSX11}.
Among them PMIA~\cite{Chen:PMIA} algorithm has stood out as the most efficient
	heuristic so far, which runs three
	orders of magnitude faster than the optimized greedy algorithm
	of~\cite{Leskovec:CELF,Chen:EIM}, while maintaining good influence spread in par with
	the greedy algorithm.

In this paper, we propose a novel scalable influence
	maximization algorithm IRIE, and demonstrate through extensive
	simulations that IRIE scales even better than PMIA, with up to two orders
	of magnitude speedup and significant savings in memory usage, while
	maintaining the same level or even better influence spread than PMIA.
We also demonstrate that while the running time of PMIA is very sensitive to
	structural properties of the network such as the clustering coefficient and the edge density, and to the cascade size, IRIE is much more stable and robust over them and always shows very fast running time.
In the greedy algorithm as well as in PMIA, each round a new seed with
	the largest marginal influence spread is selected.
To select this seed, the greedy algorithm uses Monte-Carlo simulations while
	PMIA uses more efficient local tree based heuristics to estimate
	marginal influence spread of every possible candidate.
This is especially slow for the first round where the influence spread of
	every node needs to be estimated.
Therefore, instead of estimating influence spread for each node at each round,
	we propose a novel global influence ranking method IR derived from a
	belief propagation approach, which uses a small
	number of iterations to generate a global influence ranking of the nodes and then
	select the highest ranked node as the seed.
However, the influence ranking is only good for selecting one seed.
If we use the ranking to directly select $k$ top ranked nodes as $k$ seeds,
	their influence spread may overlap with one another and not result
	in the best overall influence spread.
To overcome this shortcoming, we integrate IR with a simple influence
	estimation (IE) method, such that after one seed is selected,
	we estimate additional influence impact of this seed to each node in
	the network, which is much faster than estimating marginal influence
	for many seed candidates, and then use the results to adjust next round
	computation of influence ranking.
When combining IR and IE together, we obtain our fast IRIE algorithm.
Besides being fast, IRIE has another important advantage, which
	is its memory efficiency.
For example, PMIA needs to store data structures related to the local
	influence region of every node, and thus incurs a high memory overhead.
In constrast, IRIE mainly uses global iterative computations without storing
	extra data structures, and thus the memory overhead is small.

We conduct extensive experiments using synthetic networks as well as
	five real-world networks with size
	ranging from $29K$ to $69M$ edges, and
	different IC model parameter settings.
We compare IRIE with other state-of-the-art algorithms including
	the optimized greedy algorithm, PMIA, simulated annealing (SA) algorithm
	proposed in~\cite{JSCWSX11}, and some baseline algorithms including the PageRank.
Our results show that
	(a) for influence spread, IRIE matches the greedy
	algorithm and PMIA while being significantly better than SA and PageRank
	in a number of tests; and
	(b) for scalability, IRIE is some orders of magnitude faster than
	the greedy algorithm and PMIA and is comparable or faster than SA; and
    (c) for stability IRIE is much more stable and robust over structural properties of the network and the cascade size than PMIA and the greedy algorithm.


Moreover, to show the wide applicability of our IRIE approach, we also
	adapt IRIE to the IC-N model, which considers negative opinions emerging
	and propagating in networks~\cite{Chen:IC-N}.
Our simulation results again show that IRIE has comparable influence
	coverage while scales much better than the MIA-N heuristic
	proposed in~\cite{Chen:IC-N}.
%

\vspace{0.15cm}
\noindent{\bf Related Work.} Domingo and Richardson~\cite{Domingos:MNVC} are the first to study
influence maximization problem in probabilistic settings. Kempe et
al.~\cite{Kempe:MSI} formulate the problem of finding a subset of
influential nodes as a combinatorial optimization problem and show
that influence maximization problem is NP-hard. They propose
a greedy algorithm which guarantees $(1-1/e)$ approximation
ratio. However, their algorithm is very slow in practice and not
scalable with the network size. In~\cite{Leskovec:CELF}, \cite{Goyal:CELF++}, authors propose lazy-foward optimization that
	significantly speeds up the greedy algorithm, but it still cannot scale
	to large networks with hundreds of thousands of nodes and edges.
A number of heuristic algorithms are also proposed~\cite{Kimura:TID,Chen:EIM,Chen:PMIA,SuriN10,JSCWSX11} for
	the independent cascade model.
SPM/SP1M of~\cite{Kimura:TID} is based on shortest-path computation, and
	SPIN of~\cite{SuriN10} is based on Shapley value computation.
Both SPM/SP1M and SPIN have been shown to be not
	scalable~\cite{Chen:PMIA,CYZ10}.
Simulated annealing approach is proposed in~\cite{JSCWSX11}, which provides
	reasonable influence coverage and running time.
The best heuristic algorithm so far is believed to be the PMIA algorithm
	proposed by Chen et al.~\cite{Chen:PMIA}, which provides matching
	influence spread while running at three orders of magnitude faster than
	the optimized greedy algorithm.
PageRank~\cite{BP98} is a popular ranking algorithm for ranking web pages
	and other networked entities, and it considers diffusion
	processes whose corresponding transition matrix must have column sums
	equal to one.
Hence it can not be directly used for the influence spread estimation.
Our algorithm IR overcomes this shortcoming, and uses equations more
	directly designed for the IC model.
More importantly, our IRIE algorithm integrates influence ranking with
	influence estimation together with the greedy approach, overcoming the
	general issue of ignoring overlapping influence coverages suffered
	by all pure ranking methods.
Our simulation results also demonstrate that IRIE performs much better than
	PageRank in influence coverage.
The IC-N model is proposed in~\cite{Chen:IC-N} to consider the emergence
	and propagation of negative opinions due to product or service quality
	issues.
A corresponding MIA-N algorithm, an extension of PMIA is proposed for
	influence maximization under IC-N.
We show that our IRIE algorithm adapted to IC-N also outperforms
	MIA-N in scalability.
Recently, Goyal et al. propose a data-based approach to social influence maximization~\cite{data}. They define a new propagation probability model called \textit{credit distribution model}, which reveals how influence flows in the networks based on datasets and propose a novel algorithm for influence maximization for that model.
Scalable algorithms for a related model called linear threshold model has
	also been studied~\cite{CYZ10}.
It is a future work to see if our IRIE approach could be applied to
	further speed up scalable algorithms for the linear threshold model.

The rest of the paper is organized as follows. Section 2
	describes problem statement and preliminaries. Section 3 provides our
	\IRIE~algorithm and its extension for IC-N model. Section 4 shows
	experimental results, and Section 5 contains the conclusion.

\section{Model and Problem Setup}
\subsection{Influence Maximization Problem and IC Model}
Influence Maximization problem~\cite{Kempe:MSI} is a discrete optimization
	problem in a social network that chooses an optimal initial seed set
	of given size to maximize influence under a certain information diffusion
	model.
In this paper, we consider Independent Cascade (IC) model as the information
	diffusion process.
We first introduce IC model, then provide a formal definition of Influence
	Maximization problem under the IC model.
Let $G = (V, E)$ be a directed graph for a social network and $P_{uv} \in [0, 1]$
	be an edge propagation probability assigned to each edge $(u, v) \in E$.
Each node represents a user and each edge corresponds to a social relationship
	between a pair of users.
In the IC model, each node has either an active or inactive state and is allowed to
	change its state from inactive to active, but not the reverse direction.

Given a seed set $S$, the process of IC model is as follows : At step
	$t = 0$, all seed nodes $u \in S$ are activated and added to $S_0$.
At each step $t > 0$, a node $u \in S_{t-1}$ tries to affect its inactive
	out-neighbors $v \in N^{out}(u)$ with probability $P_{uv}$ and all the
	nodes activated at this step are added to $S_{t}$.
This process ends at a step $t$ if $|S_t| = 0$. Note that every activated
	node $u$ belongs to just one of $S_i$, where $i = 0, 1, ..., t$.
Hence, it has a single chance to activate its neighbors $v \in N^{out}(u)$ at
	the next step that it is activated.
This activation of nodes models the spread of information among people by the
	word-of-mouth effect as a result of marketing campaigns.
Under the IC model, let us define our influence function $\sigma(S)$ as
	the expected number of activated nodes given a seed set.

Formally, Influence Maximization problem is defined as follows : Given a
	directed social network $G = (V, E)$ and $P_{uv}$ for each edge $(u, v)
	\in E$, influence maximization problem is to select a seed set
	$S \subseteq V$ with $|S| = k$ that maximizes influence $\sigma(S)$ under
	the IC model.
In \cite{Kempe:MSI}, it is shown that the exact computation	of optimum solution
	for this problem is NP-hard, but the Greedy algorithm achieves $(1 - 1/e)$
	-approximation by proving the facts that the influence function $\sigma$
	is non-negative, monotone, and submodular.
A set function $f$ is called monotone if $f(S) \leq f(T)$ for all $S
	\subseteq T$, and the definition of submodular function is described at
	Definition \ref{def_submodular}.

\newtheorem{definition}{Definition}
\begin{definition}
\label{def_submodular}
A set function $ f : 2^V \rightarrow R$ is submodular if for every $S
	\subseteq T \subseteq V$ and $v \in V $, $f(S \cup u) - f(S)
	\geq f(T \cup u) - f(T)$.
\end{definition}

\newtheorem{theorem}{Theorem}
\begin{theorem}
\label{thm_approximation}
\cite{Kempe:MSI} For a non-negative, monotone, and submodular influence
	function $\sigma$, let S be a size-$k$ set obtained by the 
	greedy hill-climbing algorithm in Algorithm \ref{GREEDY}. Then $S$
	satisfies $\sigma(S) \geq (1 - 1/e) \cdot \sigma(S^*)$ where $S^*$ is
	an optimum solution.
\end{theorem}

At each step, Algorithm \ref{GREEDY} computes marginal influence of every
	node $u \in V \setminus S$ and then add the maximum one into the seed
	set $S$ until $|S| = K$.
Although the greedy algorithm guarantees constant-approximation solutions
	and is easy to implement, computing the influence function $\sigma(S)$
	is proven to be \#P-hard~\cite{Chen:PMIA}.
To estimate influence function $\sigma(S)$, Monte-Carlo simulation has been
	used in many previous works~\cite{Kempe:MSI, Leskovec:CELF, Chen:EIM,
	Goyal:CELF++}.
Although Monte-Carlo simulation provides the best accuracy among existing
	measures of influence function, the Greedy algorithm with Monte-Carlo
	simulation takes days or weeks in large networks with millions of nodes
	and edges.
Many heuristic measures have been used to estimate influence function
	such as \textit{Shortest-path computation}~\cite{Kimura:TID},
	\textit{Shapley value computation}~\cite{SuriN10},
	\textit{Effective diffusion values}~\cite{JSCWSX11},
	\textit{Degree discount}~\cite{Chen:EIM},
	\textit{Community based computation}~\cite{Wang:CGA}.
They show much faster running time than the Monte-Carlo simulation, but
	result in lower accuracy than the Greedy algorithm.
Hence, it is essential to design an algorithm that has the best trade-off
	between running time and accuracy.
In this paper, we design a scalable, and memory efficient heuristic
	algorithm balancing running time and accuracy.

\begin{algorithm}
\caption{\textbf{Greedy(K)}}
\label{GREEDY}
\begin{algorithmic}[1]
\STATE initialize $S = \emptyset$
\FOR{$i \gets 1$ \TO $K$}
\STATE select $u \gets argmax_{w \in V \setminus S}(\sigma(S \cup \lbrace w \rbrace) - \sigma(S))$
\STATE $S = S \cup \lbrace u \rbrace$
\ENDFOR
\STATE output S
\end{algorithmic}
\end{algorithm}

\subsection{IC-N Model}
We also provide a generalized version of our algorithm for Independent
	Cascade model with Negative Opinions (IC-N), which has been recently
	introduced in \cite{Chen:IC-N} to model the emergence and propagation of
	negative opinions caused by social interactions.

In the IC-N model, each node has one of three states, neutral, positive,
	and negative.
Initially, every node $u \in V \setminus S$ has neutral state and may change
	its state during the diffusion process.
We say that a node $v$ is activated at time $t$ if its state is neutral at
	time $(t - 1)$ and becomes either positive or negative at time $t$.
IC-N model has a parameter $q$ called quality factor which is a probability
	that a node	is positively activated by a positive in-neighbor.

Given a seed set $S$, the IC-N model works as follows : Initially at time
	$t = 0$, for each node $u \in S$, $u$ is activated positively with
	probability $q$ or negatively with probability $1 - q$, independently of
	all other activations.
At a step $t > 0$, for any neutral node $v$, let $A_{t}(v) \subseteq N^{in}(v)$
	be the set of in-neighbors of $v$ that are activated at step $t-1$ and
	$\pi_t(v) = $  $\langle u_1, u_2, ..., u_m\rangle$
	be a randomly permuted sequence
	of nodes $u_i$ where $u_i \in A_t(v), i = 1, 2, ..., m$.
Each node $u_i \in \pi_t(v)$ tries to activate $v$ with an independent
	probability $P_{u_iv}$ in the order of $\pi_t(v)$.
This process ends at time $t$ when there is no activated node at time $(t - 1)$.

If any node in $A_{t-1}(v)$ succeeds in activating $v$, $v$ is
	activated at step $t$ and becomes either positive or
	negative.
The state of $v$ is decided by the following rules : If $v$ is activated by a
	negative node $u$, then $v$ becomes negative. If $v$ is activated by a
	positive node, it becomes positive with probability $q$, or negative with
	probability $1 - q$.
Those rules reflect negativity bias phenomenon --- negative opinions usually
	dominate over positive opinions well known
	in social psychology~\cite{Rozin:NB}.

In the IC-N model, the influence function of a seed set $S$ in a social
	network $G$ with quality factor $q$ is defined as the expected number of
	positive nodes activated in the graph, and is denoted as
	$\sigma_G(S,q)$.
In \cite{Chen:IC-N}, Chen et al. show that $\sigma_G(S,q)$ is always
	monotone, non-negative and submodular.
Therefore, Algorithm~\ref{GREEDY} also guaranteeing $(1 - 1/e)$-approximation of an optimum solution for influence maximization
	problem under the IC-N model.


\section{Our Algorithm}
In this section, we describe our algorithms for influence
	maximization. As in the greedy algorithm and PMIA, at each round of
	\IRIE, it selects a node $u$ with the largest marginal influence
	estimate
	$\sigma(S \cup \{u\}) - \sigma(S)$. For a given seed set $S$, let
	$\sigma(u|S) = \sigma(S \cup \{u\}) - \sigma(S)$. The
	Greedy algorithm estimates $\sigma(u|S)$ by a Monte-Carlo simulation
	and PMIA generates local tree structures  for all
	$u \in V$ inducing slow running time.
The novelty of our algorithm lies in that we derive a system of linear
	equations for $\lbrace\sigma(u|S)\rbrace_{u \in V}$ whose solution can
	be computed fast by an iterative method. Then we use these computed
	values as our estimates of $\lbrace\sigma(u|S)\rbrace_{u \in V}$.

\subsection{Simple \FULLIR}
We first explain our formula for $\lbrace\sigma(u|S)\rbrace_{u \in V}$
	when $S = \emptyset$. Let $\sigma(u) = \sigma(u|\emptyset)$.
The basic idea of our formula lies in that the influence of a node $u$ is
	essentially determined by the influences of $u$'s neighbors under the IC
	model.
First suppose that graph $G = (V, E)$ is a tree graph. For $(v, u) \in E$,
	we define $m(u, v)$ to be the expected number of activated nodes when
	$S = \{u\}$	and $(u, v)$ is removed from $E$.
Note that for a tree graph $G$, $m(u, v)$ is the expected influence from $u$
	excluding the direction toward $v$. Let $\tilde{\sigma}(u)$ and 
	$\tilde{m}(u, v)$ be our estimates of $\sigma(u)$ and $m(u, v)$ respectively.
	We compute $\tilde{\sigma}(u)$ and $\tilde{m}(u, v)$ from the following formulas.
	
\begin{equation}
\label{e_gm}
\tilde{\sigma}(u) = 1 + \sum_{v \in N^{out}(u)} P_{uv} \cdot
\tilde{m}(v, u),
\end{equation}	

\begin{equation}
\label{e_m}
\tilde{m}(u, v) = 1 + \left( \sum_{w \in N^{out}(u), w \ne v} P_{uw} \cdot \tilde{m}(w, u) \right).
\end{equation}

Note that equation (\ref{e_m}) forms a system of $|E|$ linear equations on $|E|$
	variables. When $G$ is a tree, (\ref{e_m}) has a unique solution.
We prove correctness of (\ref{e_gm}) and (\ref{e_m}) by Theorem \ref{thm2}.
The proof of Theorem \ref{thm2} is described in Appendix A.

\begin{theorem}
\label{thm2}
For any tree graph, for each node $u$, $\tilde{\sigma}(u) = \sigma(u)$, and
	for each edge $(v, u) \in E$, $\tilde{m}(u, v) = m(v, u)$.
\end{theorem}
Even when $G$ is not a tree, we can define the same equations (\ref{e_gm}) and (\ref{e_m}).
In this case, the $\tilde{\sigma}(u)$ computed from \ref{e_gm}) and (\ref{e_m}) corresponds to the influence of $u$
	when we allow multiple counts of influence from $u$ to each node via different
	paths. Note that this approach has a similarity with the popular \textit{Belief
	Propagation(BP)} algorithm.
As in the BP, one natural way to compute the solution of (\ref{e_gm}) and (\ref{e_m})
	is using an iterative message passing algorithm. 

This iterative algorithm, which
	we call \FULLBP~(\BP), is described in Algorithm \ref{BP}.

\begin{algorithm}
\caption{\textbf{\FULLBP}}
\label{BP}
\begin{algorithmic}[1]
\FORALL{$(u, v) \in E$}
\STATE $\tilde{m}_0(u, v) \gets 1$
\ENDFOR
\REPEAT  
\STATE $t \gets t + 1$
\FORALL{$(v, u) \in E$}
\STATE $m_t(u, v) \gets 1 + \alpha \cdot ( \sum_{w \in N^{out}(u), w \ne v} P_{uw} \cdot \tilde{m}_{t-1}(w, u))$\label{alg:getm}
\ENDFOR
\UNTIL{$\forall(u, v) \in E$, $\tilde{m}_t(u, v) = \tilde{m}_{t-1}(u, v)$}
\FORALL{$u \in V$}
\STATE $\tilde{\sigma}(u) \gets 1 + \sum_{v \in N^{out}(u)} P_{uv} \cdot \tilde{m}_t(v, u)$
\ENDFOR
\end{algorithmic}
\end{algorithm}

Although \BP~computes good estimates of $\sigma(u)$ for tree and general
	graphs, its running time may be slow since one iteration of \BP~takes
	$O(\sum_{v \in V} d_{in}(v) \cdot d_{out}(v))$ time where $d_{in}(v)$
	and $d_{out}(v)$ is the in-degree and out-degree of $v$ respectively.
We observe that for most nodes $u$, $m(u, v)$'s are similar for any $v \in N^{in}(u)$
	since the out-degree of $u$ is not too small.
Based on this observation, by substituting the same variable $r(u)$ for all the
	$m(u, v)$, $v \in N^{in}(u)$, we obtain our formulas for the simplified
	expected influence $r(u)$ as follows :

\begin{equation}
\label{e_g}
r(u) = 1 +  \left( \sum_{v \in N^{out}(u)} P_{uv} \cdot r(v) \right).
\end{equation}

Note that equation (\ref{e_g}) forms a system of $|V|$ linear equations
	on $|V|$ variables. Let $X = \left(r(u)\right)_{u \in V}$, and the
	influence matrix $A \in \mathbb{R}^{|V|\times|V|}$ be $A_{uv} = P_{uv}$.
	Let $B = (1, 1, \ldots , 1)^T \in \mathbb{R}^{|V|}$. Then
	(\ref{e_g}) becomes	
\begin{equation}
\label{e_XA}
X = AX + B.
\end{equation}

If $\lim \limits_{k \to \infty} A^k = 0$, the solution of
	(\ref{e_XA}) becomes
\begin{displaymath}
\begin{split}
&(I - A)X = B. \\
&(I + A + A^2 + \cdots )(I - A)X = (I + A + A^2 + \cdots )B.
\end{split}
\end{displaymath}
\begin{equation}
\label{e_XB}
\therefore X = B + AB + A^2B + \cdots
\end{equation}

Note that $(A^k)_{uv}$ is the summation of the expectation of influence
	paths so that the diffusion process begins from a single node set
	$\{u\}$ and it activates a node $v$ after exactly $k$ number of
	iterations when we allow loops in the paths.
Hence $(A^k\cdot B)_u$ is equal to the expectation of \textit{relaxed}
	influence of node $u$ after exactly $k$ number of iterations where
	\textit{relaxed} means that we allow multiple counts of influence on
	some nodes and loops in the paths.
	
Hence, from (\ref{e_XB}), $X_u$ is the expectation of \textit{relaxed}
	influence of node $i$.
Note that $X_u$ is an upper bound of $\sigma(u)$ for all $u \in V$.
Here we assumed that $\lim \limits_{k \to \infty} A^k = 0$.
Note that otherwise there can appear a large spreading (constant fraction of
	nodes becomes influenced) even if the diffusion process begins from a
	single node.
It is known that in most real world information diffusion processes, such
	large spreading rarely happens.
Even when there is a large spreading, letting $X$ to be $X = B + AB +
	\cdots + A^kB$ for some $k$ is reasonable since it computes the
	$\textit{relaxed}$ influence of each node up to $k$ iterations.

Recall that $X_u$ computes \textit{relaxed} influence of node $u$.
Since we should not allow loops in the influence paths or multi-counts for
	the	computation of $\sigma(u)$, we introduce a damping factor
	$\alpha \in (0, 1)$ in our algorithm as follows.

\begin{equation}
\label{e_ga}
r(u) = 1 +  \alpha \cdot \left( \sum_{v \in N^{out}(u)} P_{uv} \cdot r(v) \right).
\end{equation}

Note that (\ref{e_ga}) is equivalent to
\begin{equation}
\label{e_XAA}
X = \alpha AX + B,
\end{equation}

and when $\lim \limits_{k \to \infty} (\alpha A)^k = 0$, the solution of
	(\ref{e_ga}) becomes
\begin{equation}
\label{e_XBAB}
X = B + \alpha AB + \alpha^2 A^2B + \alpha^3 A^3B + \cdots .
\end{equation}

For any $A \in \mathbb{R}^{|V|\times|V|}$, when
	$\alpha$ is smaller than the inverse of the largest eigenvalue of $A$,
	$\lim \limits_{k \to \infty} (\alpha A)^k = 0$. Moreover, if there is no
	large spreading in the given IC model, for all $\alpha \in (0, 1)$,
	$\lim \limits_{k \to \infty} (\alpha A)^k = 0$. Hence in those cases
	(\ref{e_XBAB}) becomes the solution of (\ref{e_ga}).
	
To compute $X$, we use an iterative computation obtained from (\ref{e_ga})
	as follows. Let $r^{(0)}(u) = 1$ for all $u \in V$, and
	$r^{(t)}(u) = 1 + \alpha \cdot \left( \sum_{v \in N^{out}(u)} P_{uv}
	\cdot r^{(t-1)}(v) \right)$ for all $u \in V$ and $ t = 1, 2, \ldots,$ .
	Then by using (\ref{e_XAA}) recursively, we have
\begin{displaymath}
(r^{(t)}(u))_{u \in V} = B + \alpha AB + (\alpha A)^2B + \cdots + (\alpha A)^tB.
\end{displaymath}

Hence $(r^{(t)}(u))_{u \in V}$ converges exponentially fast to the solution
	of (\ref{e_ga}) if $\lim \limits_{k \to \infty} (\alpha A)^k = 0$.
Even when there is a large spreading, $(r^{(k)}(u))_{u \in V}$, for some constant
	$k$ are	good estimates of $(\sigma(u))_{u \in V}$ as explained before.

The running time of	simple \IR~becomes significantly faster than \BP~since
	one iteration of simple \IR~takes $O(\sum_{v \in V} d_{out}(v))$ time.
We confirmed by experiments that accuracies of \BP~and simple \IR~are
	almost the same.
In Section 5, we show by extensive experiments that	\IR~runs much faster
	than the Greedy and PMIA, especially for large or dense networks.

One possible approach for influence maximization using simple \IR~would be
	selecting top-K	seed nodes with the highest $r(u)$.
We describe this algorithm in Algorithm \ref{IR}. 

\begin{algorithm}
\caption{\textbf{\FULLIR(K)}}
\label{IR}
\begin{algorithmic}[1]
\STATE $S \gets \{\}$
\FORALL{$u \in V$}
\STATE $r(u) \gets 1$
\ENDFOR
\REPEAT  
\FORALL{$u \in V$}
\STATE $r(u) \gets 1 + \alpha \cdot (\sum_{v \in N^{out}(u)} P_{uv} \cdot r(v))$
\ENDFOR
\UNTIL{the stopping criteria is met}
\REPEAT
\STATE $u \gets \mathop{\rm arg~max}\limits_{u \in V}(r(u)) $
\STATE $S \gets S \cup \{u\}$
\STATE $V \gets V - \{u\}$
\UNTIL{$K$ nodes are selected}

\end{algorithmic}
\end{algorithm}

However, simple \IR~can only compute the influence for individual nodes,
	and $\sigma(S) \neq \sum_{u \in S} \sigma(u)$ in general due to
	influence dependency among seed nodes.
In the next subsection, we propose \IRIE~as an extension of simple \IR~to overcome this shortcoming.

\subsection{\FULLIRIE}
In this subsection, we describe \IRIE, which performs an estimation of
	$\{\sigma(u|S)\}_{u \in V}$ for any given seed set $S$. Let $S$ be
	fixed and $AP_S(u)$ be the probability that node $u$ becomes activated
	after the diffusion process, when the seed set is S.
	Suppose that we can estimate $AP_S(u)$ by some algorithm. Many known
	algorithms including MIA and its extension PMIA, and Monte-Carlo
	simulation can be used for this	estimation. We call this part of
	our algorithm as \emph{\FULLIE~(\IE)}.

Suppose that the probability that a node $u$ becomes activated by $S$ is
	independent from activations of all other nodes. We have the following
	extension of (\ref{e_ga}) so that $\{r(u)\}_{u \in V}$ estimates
	$\{\sigma(u|S)\}_{u \in V}$.
\begin{equation}
\label{e_r}
r(u) = \left(1 - AP_S(u) \right) \cdot \left( 1 + \alpha \left( \sum_{v \in N^{out}(u)} P_{uv} \cdot r(v) \right)\right).
\end{equation}

Note that given $\{AP_S(u)\}_{u \in V}$, (\ref{e_r}) is a system
	of linear equations and is exactly same with (\ref{e_ga})
	when $S = \emptyset$. The factor $(1 - AP_S(u))$ indicates the
	probability that a node $u$ is not activated by a seed set $S$ and the
	remaining terms are the same as (\ref{e_ga}).
	
Let $D \in \mathbb{R}^{|V|\times|V|}$ be a diagonal matrix so that
	$D_{uu} = (1 - AP_S(u))$. Then for $X = (r(u))_{u \in V}$, (\ref{e_r})
	becomes $X = \alpha DAX + DB$. \IRIE~compute the solution of (\ref{e_r}) by
	an iterative computation as in the simple \IR. A pseudo-code of \IRIE
	is in Algorithm \ref{IRIE}. As in the simple \IR,
	when $\lim \limits_{k \to \infty} (\alpha DA)^k = 0$, the iterative
	computation of $r(u)$ converges to the solution of (\ref{e_XAA}) exponentially
	fast.
As in the simple \IR, repeating line~\ref{alg:getru} of Algorithm \ref{IRIE} for constantly
	many times computes $\{r(u)\}$ which is a good estimate of
	$\{\sigma(u|S)\}_{u \in V}$.
	
\begin{algorithm}
\caption{\textbf{\IRIE(K)}}
\label{IRIE}
\begin{algorithmic}[1]
\STATE $S \gets \{\}$
\FORALL{$u \in V$}
\STATE $r(u) \gets 1$
\STATE $AP_S(u) \gets 0$
\ENDFOR

\REPEAT
\STATE $\forall u \in S$, $AP_S(u) = 1$
\STATE $\forall u \in V \setminus S$, estimate $AP_S(u)$

\REPEAT
\FORALL{$u \in V$}
\STATE $r(u) \gets (1 - AP_S(u)) \cdot ( 1 + \alpha \cdot ( \sum_{v \in N^{out}(u)} P_{uv} \cdot r(v)))$\label{alg:getru}
\ENDFOR
\UNTIL{the stopping criteria is met}
\STATE $u \gets \mathop{\rm arg~max}\limits_{u \in V}(r(u))$
\STATE $S \gets S \cup \{u\}$
\STATE $V \gets V - \{u\}$
\UNTIL{$K$ nodes are selected}

\end{algorithmic}
\end{algorithm}

Now we explain how we estimate $AP_S(u)$. Given a seed set $S$, we compute
	the Maximum Influence Out-Aborescence (MIOA)~\cite{Chen:PMIA} of
	$s$ for all $s \in S$. MIOA is a tree-based approximation of local
	influence region of an individual $s$, assuming the influence from a
	seed node $s$ to other nodes is propagated mainly along a single path
	which gives the highest activation probability. By generating MIOA
	structure for all the seed node $s \in S$, we estimate $AP_S(u)$
	according to following equation.

\begin{displaymath}
AP_S(u) = \sum_{s \in S} AP_s(u).
\end{displaymath}
Although the equation for $AP_S(u)$ is not the exact activation
	probability from a seed set $S$ to a node $u$, simple summation
	over the activation	probability for each seed node has advantages
	in terms of of running time and memory usage while achieving very high
	accuracy as shown by experiments in Section 5. Note that the \IE~part
	can be replaced with any other algorithm that estimates $AP_S(u)$,
	making our \IRIE~algorithm to be a general framework.
	
Regarding the choice of $\alpha$, we found by extensive experiments that
	the accuracy of \IRIE~is quite similar for broad range of
	$\alpha \in [0.3, 0.9]$ for most cases.
We suggest a fixed $\alpha = 0.7$ since the \IRIE~shows almost highest
	accuracy when $\alpha = 0.7$ for most cases of our experiments.
	
\subsection{Algorithm for IC-N model}
In this subsection, we describe the extension of \IRIE~to the IC-N model,
	which we call IRIE-N.
For the IC-N model, we generalize a net influence function of a seed set $S$ as $\sigma_{net}(S) = \sigma_P(S) - \lambda \cdot \sigma_N(S)$, where $\lambda\ge 0$.
We propose a system of linear equations that estimates the net influence $\sigma_{net}(S)$ of a seed set $S$ for any $\lambda\ge 0$ under the IC-N model.

For the IC-N model, we define $AP_S(u)$ as the probability that a node $u$
	has either a positive or a negative opinion after the diffusion process
	with the seed set $S$.
In the IC-N model, note that $P_{uv}$ is the same for the positive opinion
	activation and the negative opinion activation.
Hence, if we merge the two opinions of a node into one \textit{activated}
	state, the diffusion process under the IC-N model is exactly the same as
	the IC model with the same $\{P_{uv}\}$.
So $AP_S(u)$ for the IC-N model is equal to that for the corresponding IC
	model.
Therefore, $\{AP_S(u)\}_{u \in V}$ can be computed by the same algorithm for
	the corresponding IC model.
	
The basic framework of \IRIE-N~is the same as the \IRIE. \IRIE-N~consists
	of $K$ rounds, and at each round, it selects a node $u$ with the largest
	marginal net influence $\sigma_{net}(S \cup \{u\}) - \sigma_{net}(S)$.
	Let $\sigma_P(u|S) = \sigma_P(S \cup \{u\}) - \sigma_P(S)$ and
	$\sigma_N(u|S) = \sigma_N(S \cup \{u\}) - \sigma_N(S)$. To estimate
	$\sigma_{net}(S \cup \{u\}) - \sigma_{net}(S)$, we consider
	$\sigma_P(u|S)$ and $\sigma_N(u|S)$ separately, and obtain formulas
	among them.
	
\begin{algorithm}
\caption{\textbf{\IRIE-N(K, $\lambda$)}}
\label{IRIE-N}
\begin{algorithmic}[1]
\STATE $S \gets \{\}$
\FORALL{$u \in V$}
\STATE $AP_S(u) \gets 0$, $g^P(u) \gets q$, $g^N(u) \gets 1-q$, $h(u) \gets 1$
\ENDFOR

\REPEAT
\STATE $\forall u \in S$, $AP_S(u) = 1$
\STATE $\forall u \in V \setminus S$, estimate $AP_S(u)$

\REPEAT  
\FORALL{$u \in V$}
\STATE $g^P(u) \gets (1 - AP_S(u)) \cdot q \cdot (1 + \alpha \cdot (\sum_{v \in N^{out}(u)} P_{uv} \cdot g^P(v)))$
\STATE $g^N(u) \gets (1 - AP_S(u)) \cdot ((1 - q) + \alpha \cdot (\sum_{v \in N^{out}(u)} P_{uv} \cdot ((1 - q) \cdot h(v) + q \cdot g^N(v))))$
\STATE $h(u) \gets (1 - AP_S(u)) \cdot (1 + \alpha \cdot (\sum_{v \in N^{out}(u)} P_{uv} \cdot h(v)))$
\ENDFOR
\UNTIL{the stopping criteria is met}
\STATE $u \gets \mathop{\rm arg~max}\limits_{u \in V}(g^P(u) - \lambda \cdot g^N(u)) $
\STATE $S \gets S \cup \{u\}$
\STATE $V \gets V - \{u\}$
\UNTIL{$K$ nodes are selected}

\end{algorithmic}
\end{algorithm}
	
Let $S$ be fixed. We denote $g^P(u)$ and $g^N(u)$ to be our estimates of
	$\sigma_P(u|S)$ and $\sigma_N(u|S)$ respectively. Let $h(u)$ denote
	our estimate of marginal negative influence when $u$ is activated by
	a negative activation trial. We obtain the following formulas for
	$g^P(u)$, $g^N(u)$, and $h(u)$.
	
\begin{equation}
\begin{split}
\label{e_gp}
g^P(u) = \left( 1 - AP_S(u) \right) \cdot q \cdot \left( 1 + \alpha \left( \sum_{v \in N^{out}(u)} P_{uv} \cdot g^P(v) \right) \right),
\end{split}
\end{equation}

\begin{equation}
\begin{split}
\label{e_gn}
&g^N(u) = \left( 1 - AP_S(u) \right) \cdot \\
&\left( (1 - q) + \alpha \left( \sum_{v \in N^{out}(u)} P_{uv} \cdot \left( (1 - q) \cdot h(v) + q \cdot g^N(v) \right) \right) \right),
\end{split}
\end{equation}

\begin{equation}
\label{e_h}
h(u) = \left( 1 - AP_S(u) \right) \cdot \left( 1 + \alpha \left( \sum_{v \in N^{out}(u)} P_{uv} \cdot h(v) \right) \right).
\end{equation}

In (\ref{e_gp}), $g^P(u)$ has a factor $q$ which is the probability that
	$u$ has a positive state when $u$ is chosen as a seed or $u$ is
	positively activated by one of its neighbors. In (\ref{e_gn}), $g^N(u)$
	computation considers both cases when $u$ becomes a positive state,
and $u$ becomes a negative state after a positive neighbor activates $u$.
	Equation (\ref{e_h}) has the same form as (\ref{e_r}) since nodes that
	have negative opinion only negatively activates its neighbors.
	
We compute the solution of (\ref{e_gp}), (\ref{e_gn}), and (\ref{e_h}) by
	a similar iterative computation as in the \IRIE. The pseudo-code is
	described in Algorithm \ref{IRIE-N}. Note that if the corresponding influence
	matrix $A$ satisfies that $\lim \limits_{k \to \infty} (\alpha DA)^k = 0$, the
	iterative computations of \IRIE-N~also converge exponentially fast to the 
	solution of (\ref{e_gp}), (\ref{e_gn}), and (\ref{e_h}) for any 
	$q \in [0, 1]$, and $\{AP_S(u)\}_{u \in V}$.
\section{Experiments}
We conduct extensive experiments on a number of algorithms including
	\IRIE~algorithm and other state-of-the-art algorithms for influence maximization
	on various real-world social networks.
Our experiments consider following major issues : scalability, sensitivity to propagation models, influence
	spreads, running time, and memory efficiency.

\subsection{Experimental Setup}
\subsubsection{Datasets}
We perform experiments on five real-world social networks, whose edge sizes
	range from 29K to 69M.
First, we have two (undirected) co-authorship network, collected from ArXiv General Relativity
	~\cite{Leskovec:SNAP} and DBLP Computer Science Biblography Database, denoted
	by ArXiv and DBLP respectively.
Nodes corresponds to users and edges are established by co-authorship
	among users.
We also have three (directed) friendship networks collected from Epinions.com~\cite{Leskovec:SNAP},
	Slashdot.com~\cite{Leskovec:SNAP}, and LiveJournal.com~\cite{Leskovec:SNAP},
	denoted by Epinions, Slashdot, and LiveJournal respectively.
A node corresponds to a user and a directed edge represents a trust
	relationship between users.
\begin{table}
\centering
\caption{Summary of Real-world Social Networks} \label{tab:datasets}
\begin{tabular}{|c|c|c|c|} \hline
Dataset & \#nodes & \#edges & direction \\ \hline
ArXiv & 5K & 29K & undirected\\ \hline
Epinions & 76K & 509K & directed\\ \hline
Slashdot & 77K & 905K & directed\\ \hline
DBLP & 655K & 2M & undirected \\ \hline
LiveJournal & 4.8M & 69M & directed\\
\hline\end{tabular}
\end{table}
We note that in Epinions and Slashdot, nodes are more densely connected than
	co-aurhorship networks,
	although the number of nodes for both networks are of moderate size.	
The five real-world social network datasets are summarized in
	Table~\ref{tab:datasets}.
For the scalability test, we use synthetic power-law random networks
	with various sizes generated by PYTHON web Graph Library.

\subsubsection{Propagation Probability Models}
We use two propagation probability models, the Weighted cascade (WC) model and
	the Trivalency (TR) model which have been used as standard benchmarks in previous
	works so that we can compare IRIE with previous works easily.
	
\begin{itemize}
\item{\textbf{Weighted cascade model}. Weighted cascade model proposed in
	\cite{Kempe:MSI} assigns a propagation probability to each edge by
	$P_{uv} = 1 / d_v$ where $d_v$ is the in-degree of $v$. This model can be used to explain information spreading in social networks
where the receivers of information adopts similar amount of information regardless of her indegree. For example, consider the case when everyone
reads similar number of tweets per a day in Twitter.}

\item{\textbf{Trivalency model.} Trivalency model proposed in
	\cite{Chen:PMIA} assigns a randomly selected probability from \{0.l,
	0.01, 0.001\} to each directed edge. This model represents the case when there several types of personal relations (three types in this case), and the edge propagation probability depends on the type of the relation.}
\end{itemize}

\subsubsection{Algorithms and Parameter Settings}

\begin{figure}
\centering
\numberwithin{figure}{section}
\begin{center}
\subfigure[Weighted Cascade]{\label{fig:scaleNodeWC}\includegraphics[width = 0.35 \textwidth]{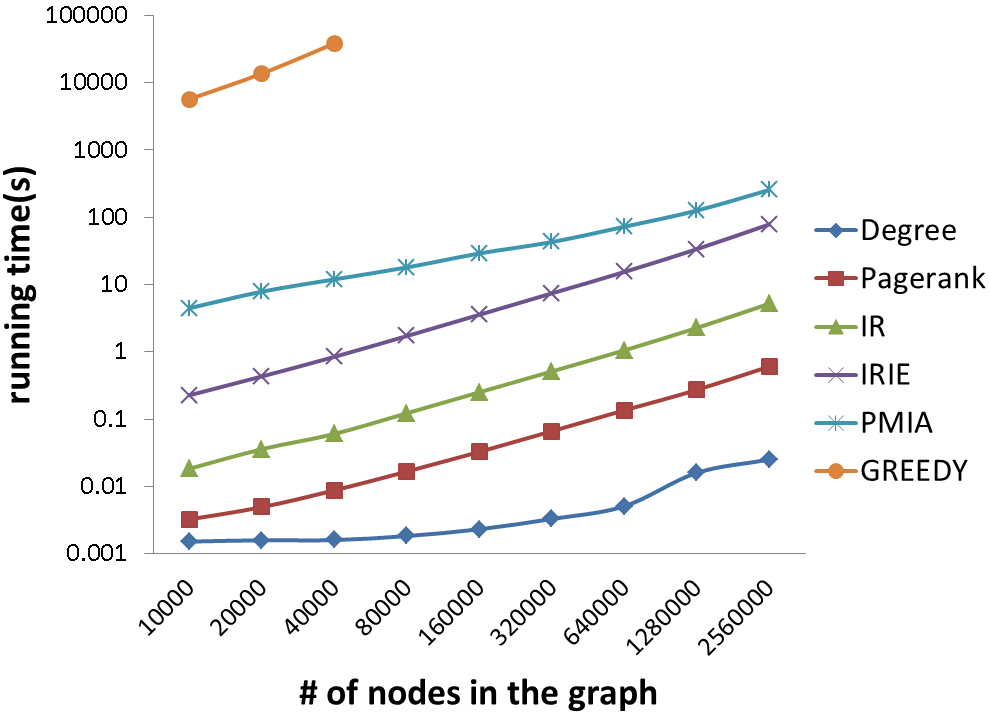}}
\subfigure[Trivalency]{\label{fig:scaleNodeTR}\includegraphics[width = 0.35 \textwidth]{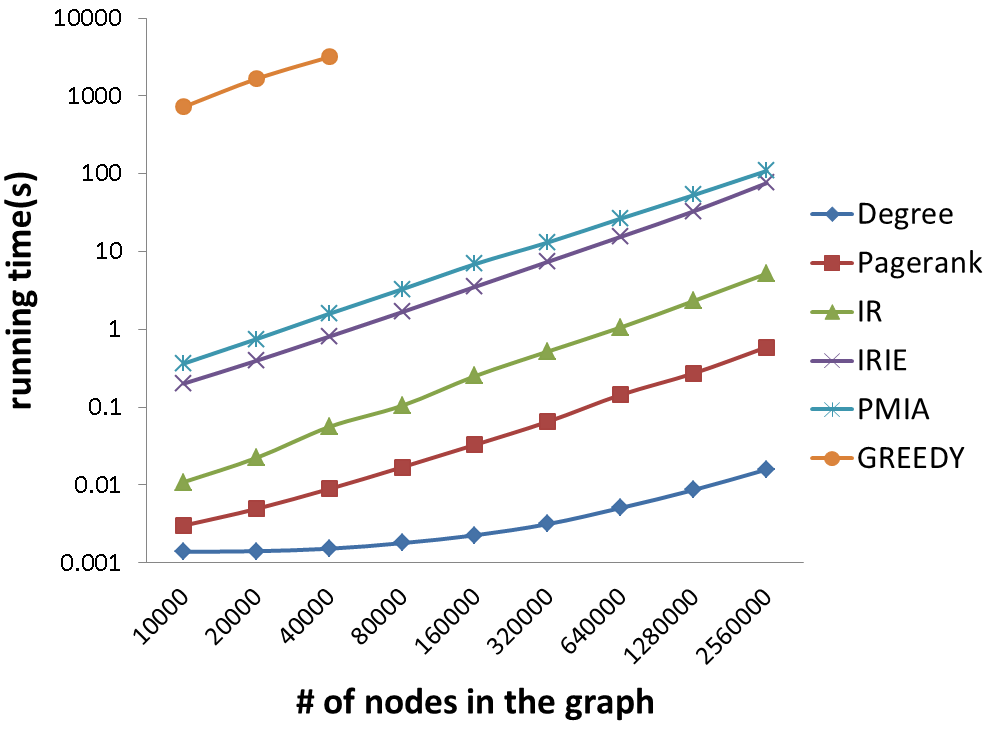}}
\\
\subfigure[Weighted Cascade]{\label{fig:scaleEdgeWC}\includegraphics[width = 0.35 \textwidth]{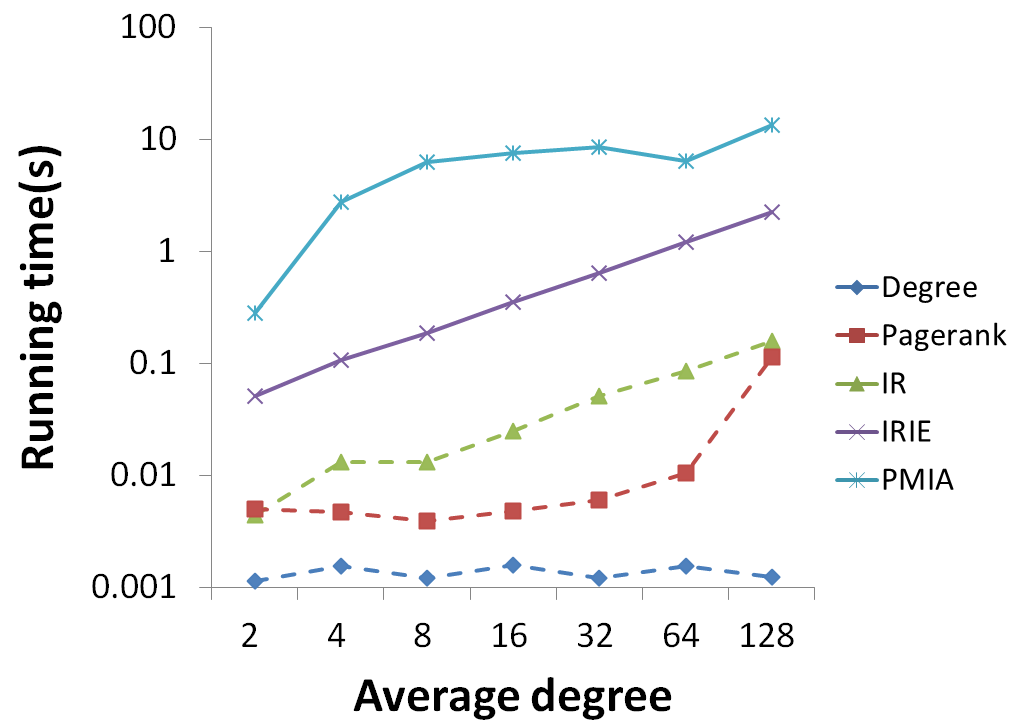}}
\subfigure[Trivalency]{\label{fig:scaleEdgeTR}\includegraphics[width = 0.35 \textwidth]{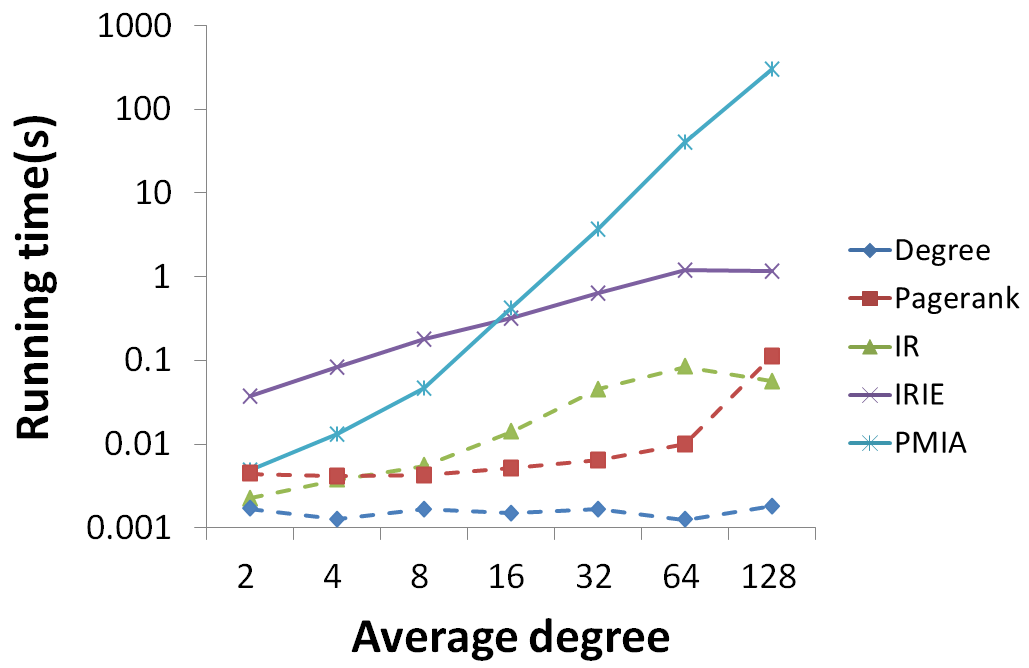}}
\caption{Scalability test for the synthetic dataset}
\label{fig:scale}
\end{center}
\end{figure}

We compare our algorithms with state-of-the-art algorithms.
The list of algorithms and corresponding parameter settings are as follows.

\textbf{Degree} A baseline algorithm selecting $K$ seed nodes with highest
	degree.

\textbf{PageRank} A baseline algorithm selecting $K$ seed nodes with highest
	ranking according to a diffusion process.
In our experiments, we used the following weighted version of
	PageRank~\cite{Chen:PMIA}.
The transition probability $TP_{uv}$ along edge $(u, v)$ is defined by
	$TP_{uv} = P_{vu} / \sum_{w \in N^{in}(u)} P_{wu}$.
The more activation probability along the edge $(u, v)$, the more transition
	probability of moving from $u$ to $v$.
We set the random jump factor of PageRank as 0.15 as in \cite{Chen:PMIA}.

\textbf{CELF} Greedy algorithm with Cost-Effective Lazy Forward(CELF)
	optimization~\cite{Leskovec:CELF}.

\textbf{SAEDV} Simulated Annealing with Effective Diffusion
	Values(SAEDV)~\cite{JSCWSX11} uses an efficient	heuristic measure
	to estimate influence of a set
	of nodes, which significantly running time of the algorithm.
We do simulations with the proposed parameter settings, as well as our
	tuned parameters for our datasets.
In our tuned parameters, we set initial temperature $T_0 = 5|V|$.
The parameters $q$ and $\bigtriangleup T$ are set as 1000 and 2000
	respectively as in~\cite{JSCWSX11}.
We use the down-hill probability to be $exp\left(\frac{\bigtriangleup f
	\cdot C_i}{\sqrt{T_t}}\right)$ where $C_i$ is the number of iterations.
We present results with better accuracy among the original parameters
	and our tuned parameters for each dataset.

\textbf{PMIA} PMIA~\cite{Chen:PMIA} restricts the influence estimation for
	a set of nodes on local shortest-paths.
The parameter $\theta $ of PMIA is set to 1/320 as in \cite{Chen:PMIA}.

\textbf{\IR} Our Algorithm \ref{IR} with $\alpha = 0.7$.

\textbf{\IRIE} Our Algorithm \ref{IRIE} with $\alpha = 0.7$. Another parameter
	$\theta$ for generating MIOA~\cite{Chen:PMIA} is set to 1/320 as in
	\cite{Chen:PMIA}.
	
As the \textbf{stopping criteria} of \IR~and \IRIE, we use the followings.
For \IR~and the first round of \IRIE, i.e., when $S = \emptyset$, we stop
	iterative computations for corresponding formulas when
	for each $u \in V$, difference between current $r(u)$ and the previous
	$r(u)$ is less than $0.0001$. Otherwise iterative computations
	run 20 rounds.
For the subsequent rounds of IRIE, we initialize each $r(u)$ by the
	output of the previous round. Since those initial values make
	the iteration converge much faster, we run the iterations of line 10-12
	of Algorithm \ref{IRIE} at most 5 times and apply the same
	stopping criteria as in the first round.

\vspace{0.15cm}
\noindent{\bf Algorithms for IC-N Model.}

\textbf{CELF-N} Greedy algorithm with cost-effective lazy forward
	optimization~\cite{Leskovec:CELF} with the influence function
	$\sigma_G(S, q)$.

\textbf{MIA-N} MIA-N proposed in \cite{Chen:IC-N} is a variation of PMIA
	for IC-N model. The parameter $\theta $ of MIA-N is set
	to 1/160 as in \cite{Chen:IC-N}.

\textbf{\IRIE-N} Our Algorithm~\ref{IRIE-N} with $\alpha$ = 0.7.
We set the parameter $\theta$ for generating MIOA~\cite{Chen:PMIA} as 1/160.
The same stopping criteria as in \IRIE~is used for \IRIE-N.

To compare the amount of influence spread of above algorithms, we run
	the	Monte-Carlo simulation on both IC and IC-N models 10,000 times
	for each seed set and take the average of the influence spreads.
Our experimental environment is a server with 2.8GHZ Quad-Core Intel i7 930 and 24GB memory.

\subsection{Experimental Results}
\subsubsection{Scalability Test for the Synthetic Dataset}
Figure \ref{fig:scale} shows the experimental results on scalability of the
	algorithms.
For Figures \ref{fig:scale}(a) and \ref{fig:scale}(b), we generate
	synthetic power-law random network datasets by increasing the number of
	nodes $|V|$ = 2K, 4K, 8K, ..., 256K while fixing the average degree
	= 10.
For Figures \ref{fig:scale}(c) and \ref{fig:scale}(d), we generate
	second synthetic power-law networks by fixing $|V| = 2K$ and increasing
	the number of edges $|E|$ = 2K, 4K, ..., 128K.
We set $K=50$, and the figures are plotted in log-log scale.
In Figures \ref{fig:scale}(a) and \ref{fig:scale}(b), \IR~and \IRIE~
	show efficient running time and scalability. PMIA is also scalable in
	the number of nodes but about 2-10 times slower than \IR~and \IRIE.
In \ref{fig:scale}(c) and \ref{fig:scale}(d), \IR~and \IRIE~shows much
	better running time and scalability over the average degree than
	PMIA.
Hence we find that \IR~and \IRIE~show much more robust performance over the
	edge density than PMIA in terms of scalability.

\subsubsection{Sensitivity to Propagation Probability Models}
\begin{figure}
\centering
\numberwithin{figure}{section}
\includegraphics[width = 0.45 \textwidth]{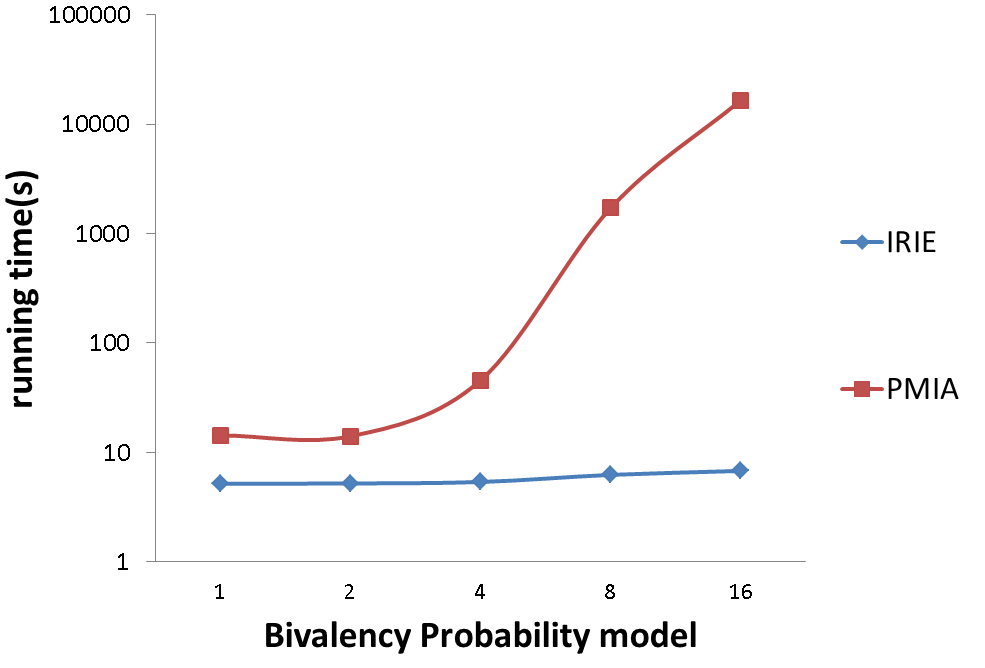}
\caption{Sensitivity of algorithms under various Bivalency models for
	the Epinions dataset}
\label{fig:sensitivity}
\end{figure}
\begin{figure*}
\centering
\numberwithin{figure}{section}
\begin{center}
\subfigure[Arxiv-WC]{\label{fig:ICArXivWC}\includegraphics[width = 0.35 \textwidth]{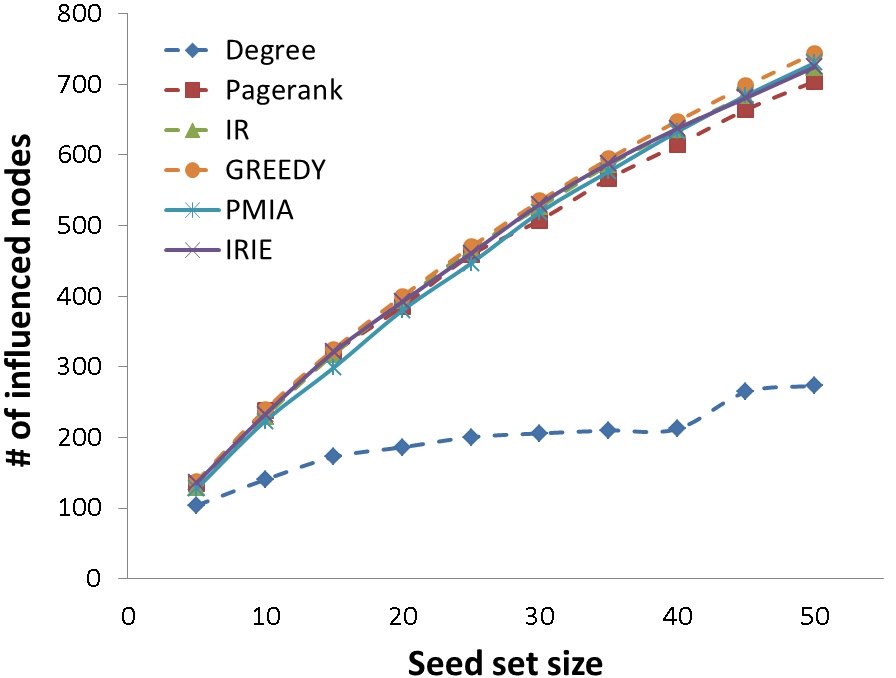}}
\subfigure[Arxiv-TR]{\label{fig:ICArXivTR}\includegraphics[width = 0.35 \textwidth]{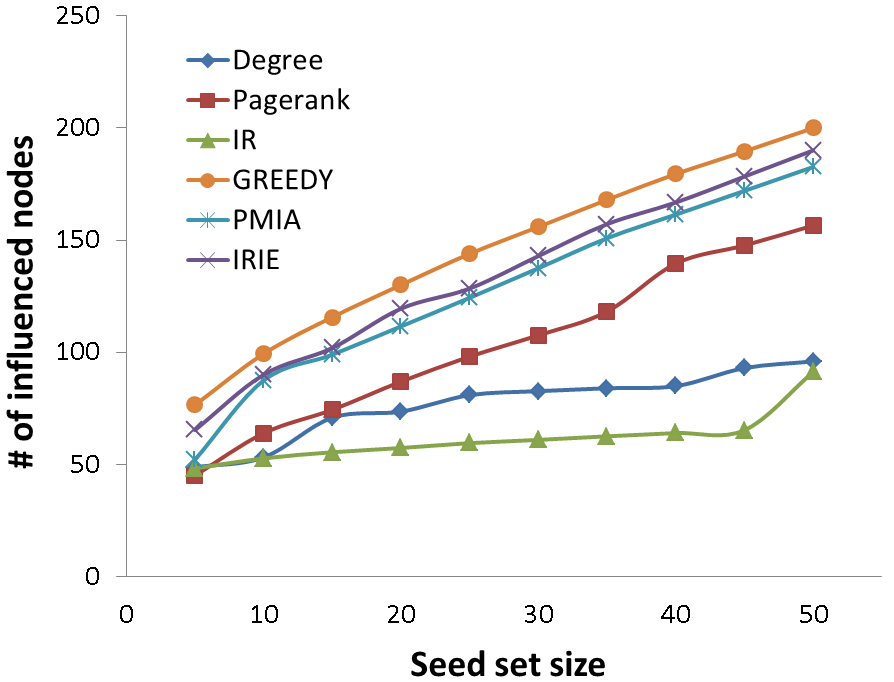}}
\\
\subfigure[Epinions-WC]{\label{fig:ICEpinionsWC}\includegraphics[width = 0.35 \textwidth]{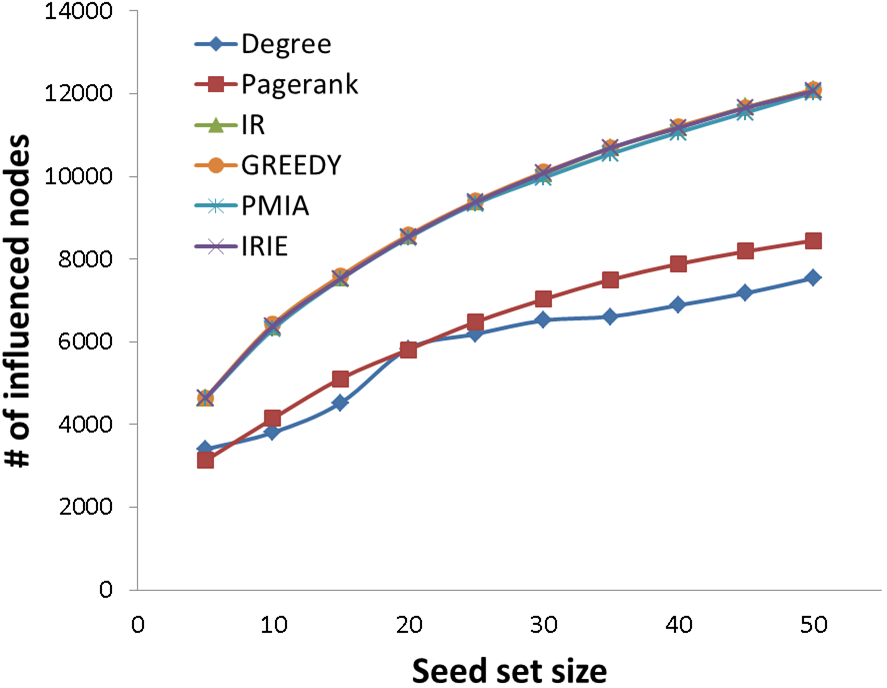}}
\subfigure[Epinions-TR]{\label{fig:ICEpinionsTR}\includegraphics[width = 0.35 \textwidth]{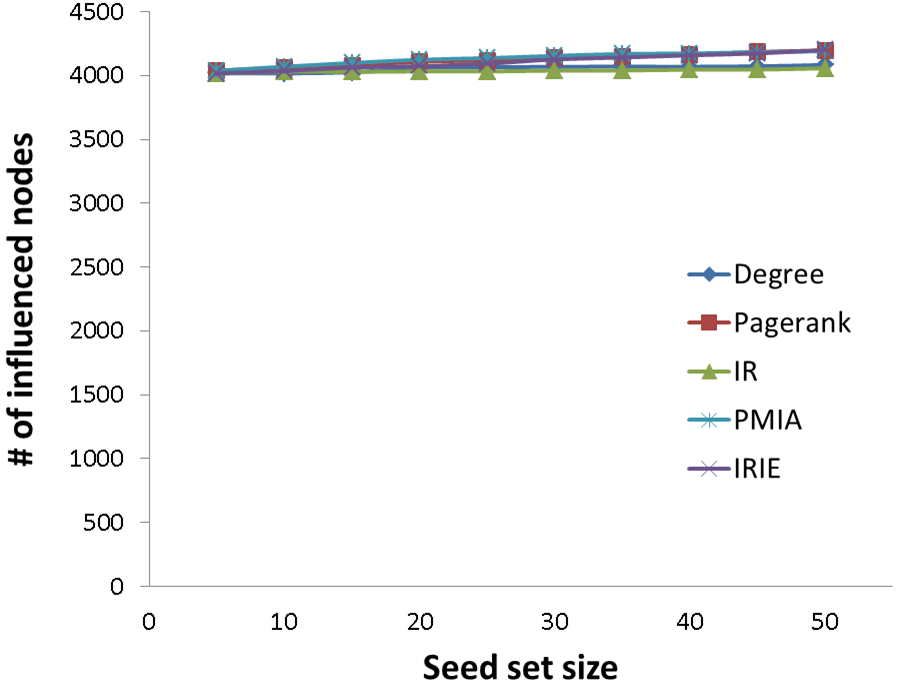}}
\\
\subfigure[Slashdot-WC]{\label{fig:ICSlashdotWC}\includegraphics[width = 0.35 \textwidth]{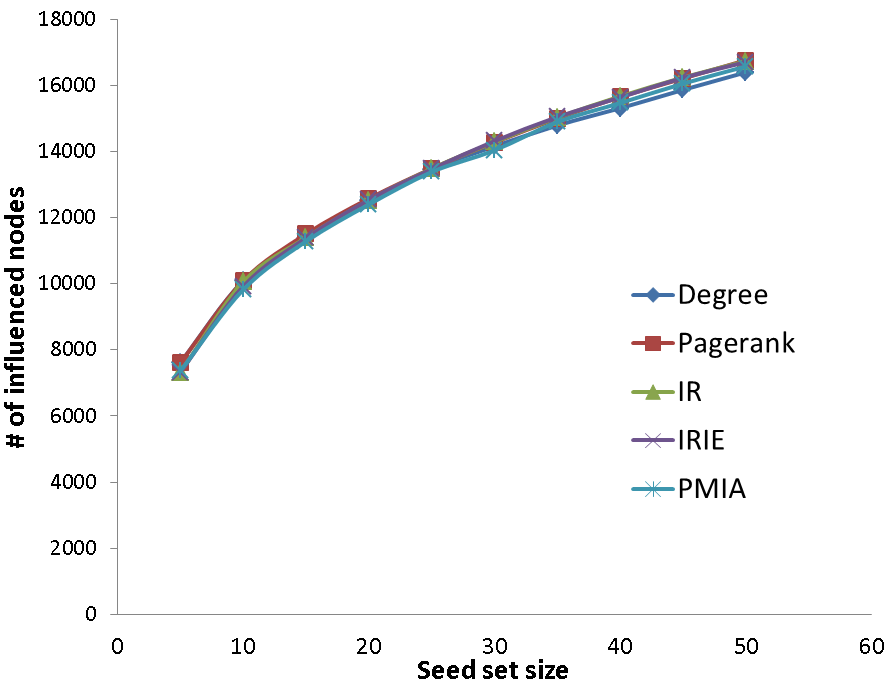}}
\subfigure[Slashdot-TR]{\label{fig:ICSlashdotTR}\includegraphics[width = 0.35 \textwidth]{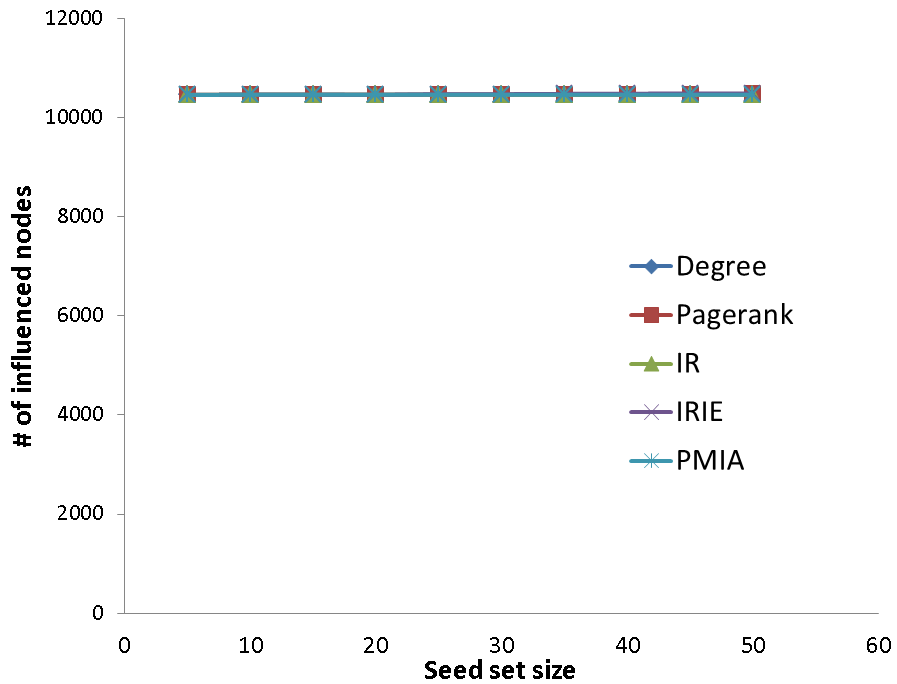}}
\\
\subfigure[DBLP-WC]{\label{fig:ICDBLPWC}\includegraphics[width = 0.35 \textwidth]{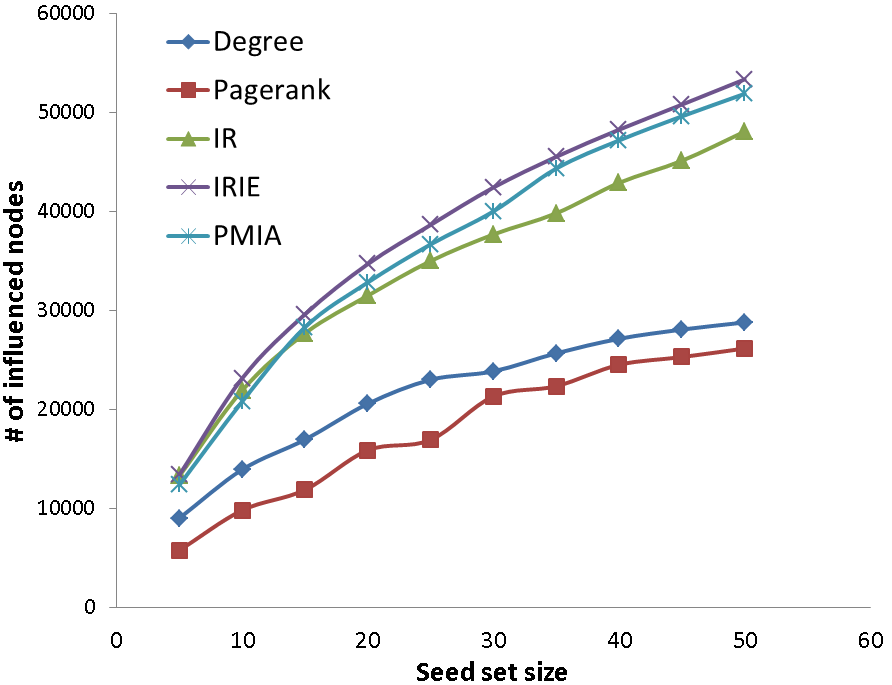}}
\subfigure[DBLP-TR]{\label{fig:ICDBLPTR}\includegraphics[width = 0.35 \textwidth]{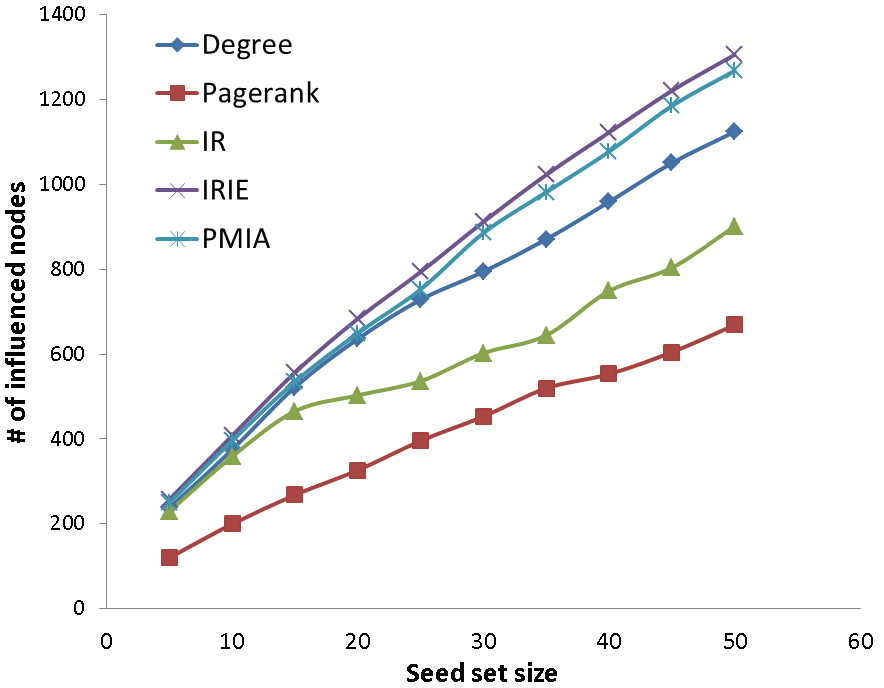}}
\end{center}
\caption{Influence spreads for IC model}
\label{fig:IC}
\end{figure*}
We compare \IRIE~with PMIA in terms of the sensitivity of running time to
	propagation probability models.
In this experiment, we compare running times of \IRIE~and PMIA on Epinions
	dataset for various bivalency models described as follows.
For each propagation model indexed by $i\in \{1,2,4,8,16\}$, edge 	
	propagation probabilities are randomly assigned from $i~\times~\{0.01,
	0.001\}$.
We set the seed size $K=50$. In Figure~\ref{fig:sensitivity},
	\IRIE~shows much faster and more stable running time than PMIA.
The running time of \IRIE~slightly increases as the edge probability
	increases, while the running time of PMIA increases dramatically around
	$i=8$, where the spread size becomes large.
Especially, \IRIE~is more than 1000 times faster than PMIA for the (0.16,
	0.016)-bivalency model. Hence we observe that while the running time of
	PMIA is quite dependent on the spread size and the propagation model,
	the running time of \IRIE~ is very stable over them.


\subsubsection{Influence Spread for the Real-World Datasets}
We compare influence spread for each algorithms on the five real-world datasets.
The seed size $K$ is set from 1 to 50 to compare the accuracies of algorithm in various range of seed sizes.
Figure \ref{fig:IC} (a)-(h) and Table~\ref{tab:infLiveJournal}
	show the experimental results on influence spread.
We run the CELF only for Arxiv, and Epinions(for the WC) since CELF runs too
	long for other datasets.
We did Monte-Carlo simulation for LiveJournal only for $K$=50 since it takes
	too long time.

In general, CELF performs almost the best influence spread for both the WC and the TR models.
However, \IRIE~shows almost similar performance with CELF in all cases.
PMIA also shows high performance but 1-5\% less influence spread than \IRIE~for all cases except for the Epinions TR.
\IR~shows high performances for the WC models, but not quite good in the TR models.
Hence we observe that \IE~part of \IRIE~is necessary to achieve robust performance in various steps.
The baseline algorithms Degree and PageRank show low Performances for many cases such as Arxiv, Epinions, and DBLP.
Unlike the Greedy based approaches, SAEDV computes the seed set for each $K$
independently.
\begin{table}
\centering
\caption{Influence spread at 50-seed set for LiveJournal}
	\label{tab:infLiveJournal}
\begin{tabular}{|c|c|c|} \hline
Algorithm & Weighted Cascade & Trivalency \\ \hline
IRIE & 74830.5 & 629694 \\ \hline
IR & 75861.2 & 629484 \\ \hline
PMIA & 71566.5 & 629512  \\ \hline
PageRank & 51162.6 & 629892 \\ \hline
Degree & 52162.3 & 629498\\ \hline
\end{tabular}
\end{table}
\begin{table}
\centering
\caption{Influence spread at 50-seed set of SAEDV and \IRIE}
\label{tab:sa}
\begin{tabular}{c|c|c|c|c|} \cline{2-5}
& \multicolumn{2}{|c|}{WC} & \multicolumn{2}{|c|}{TR} \\ \hline
\multicolumn{1}{|c|}{Dataset} & SAEDV & \IRIE & SAEDV & \IRIE \\ \hline
\multicolumn{1}{|c|}{ArXiv} & 669.755 & 724.666 & 185.369 & 190.006\\ \hline
\multicolumn{1}{|c|}{Epinions} & 11177.3 & 12063 & 4176.35 & 4200.22\\ \hline
\multicolumn{1}{|c|}{Slashdot} & 14803 & 16712.3 & 10467.7 & 10490.2\\ \hline
\multicolumn{1}{|c|}{Amazon} & 487.671 & 824.795 & 79.5405 & 82.041\\ \hline
\multicolumn{1}{|c|}{DBLP} & 33730 & 53334.8 & 1175.53 & 1304.81\\ \hline
\end{tabular}
\end{table}
Hence we include Table~\ref{tab:sa} that shows the influence spread comparison of SAEDV with \IRIE~for $K$=50.
Table~\ref{tab:sa} clearly shows that \IRIE~outperforms SAEDV in terms of influence spread by large margin for most cases.
Hence we conclude that \IRIE~shows very high accuracy and robustness in most environments.

\subsubsection{Running Time and Memory Usage for the Real-World Datasets}
\begin{figure}
\centering
\numberwithin{figure}{section}
\includegraphics[width = 0.45 \textwidth]{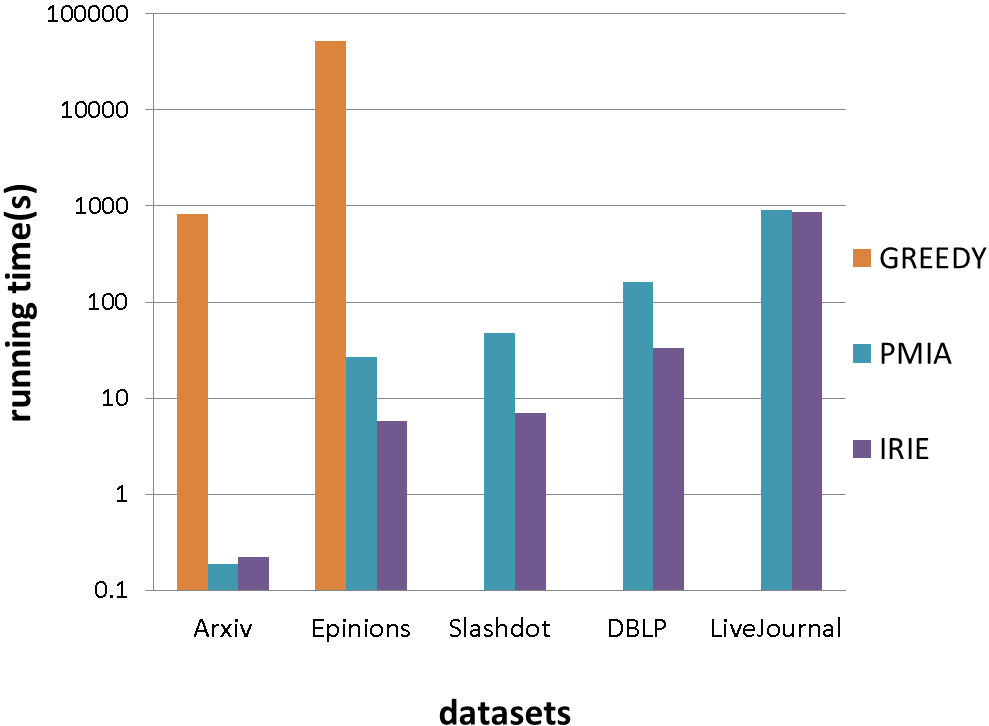}
\includegraphics[width = 0.45 \textwidth]{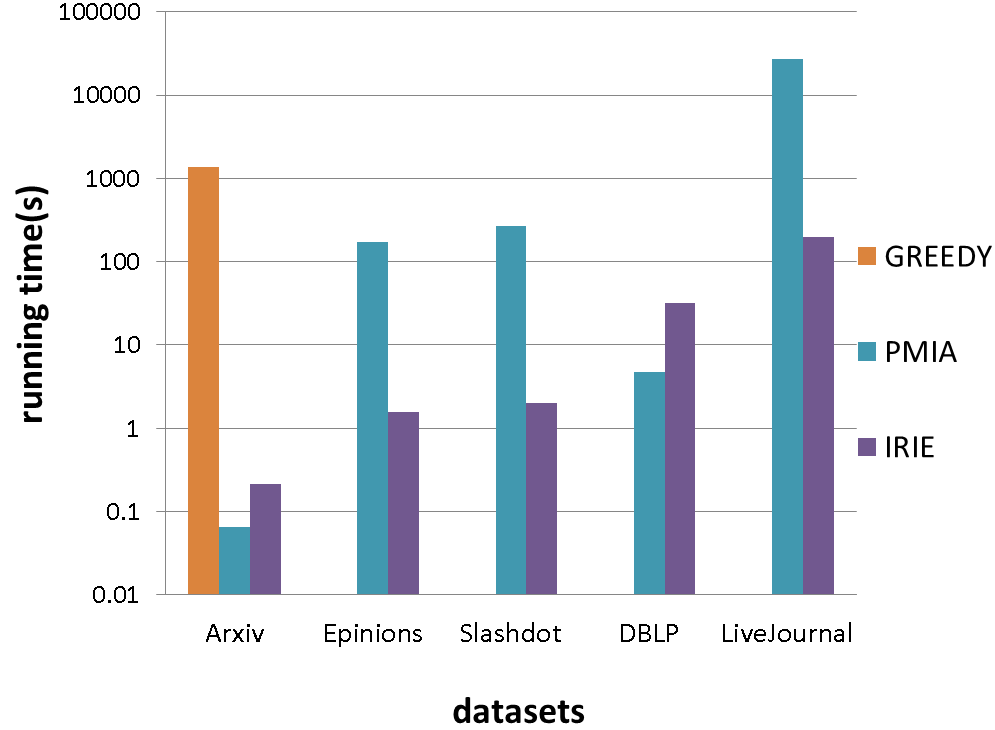}
\caption{Running time of algorithms under IC model}
\label{fig:running}
\end{figure}
\begin{table*}
\centering
\caption{Memory usages of IRIE and PMIA}\label{tab:memory}
\begin{tabular}{c|c|c|c|c|c|c|} \cline{2-7}
& \multicolumn{3}{|c|}{WC} & \multicolumn{3}{|c|}{TR} \\ \hline
\multicolumn{1}{|c|}{Dataset} & File size & PMIA & IRIE & File size & PMIA & IRIE\\ \hline
\multicolumn{1}{|c|}{ArXiv} & 715KB & 14MB & 8.7MB & 582KB & 10MB & 8.7MB\\ \hline
\multicolumn{1}{|c|}{Epinions} & 18MB & 135MB & 35MB & 15MB & 143MB & 35MB\\ \hline
\multicolumn{1}{|c|}{Slashdot} & 24MB	& 280MB	& 39MB & 19MB & 340MB & 40MB\\ \hline
\multicolumn{1}{|c|}{DBLP} & 88MB &	1.1GB &	160MB & 82MB & 357MB & 158MB\\ \hline
\multicolumn{1}{|c|}{LiveJournal}	& 2.4GB & 10.1GB & 3GB & 2GB & 16GB & 3GB\\
\hline\end{tabular}
\end{table*}

We also checked the running time of the algorithms on the real-world social
	networks.
Figure \ref{fig:running} shows the results.
The left and right figures in \ref{fig:running} corresponds to the WC model
	and the TR model respectively.
In each figure, datasets are aligned in increasing order of network sizes
	from left to right.
For both the WC and the TR model, \IRIE~is more than 1000 times faster than
	the CELF.
Also in most cases, \IRIE~is quite faster than PMIA.
Note that the running time of \IRIE~is increasing as the dataset size
	increases from Arxiv to LiveJournal.
However, the running times of PMIA are somewhat unstable, resulting in
	longer running times even in smaller graph in both the numbers of nodes
	and edges.
	
Note that although the numbers of nodes and edges of Epinions and Slashdot
	are smaller than those of DBLP,	the running times of PMIA for Epinions
	and Slashdot are much larger than for DBLP.
One possible explanation is that the running time of PMIA is sensitive to
	structural properties of the network such as the clustering coefficient
	(Epinions and Slashdot are social network dataset which contains many
	triangles) and edge density, and the spread size (note that Epinions TR
	and Slashdot TR induce larger spread than DBLP TR) which matches the
	results of the scalability test and the sensitivity test in Sections
	4.2.1 and 4.2.2.
Hence, we conclude that \IRIE~shows much more stable and faster running time
	than PMIA in various networks.
	
Table~\ref{tab:memory}
	shows the experimental results on the amount of memory used by
	algorithms for the WC and the TR model respectively.
In the table, file sizes indicate the size of raw text data files, and PMIA
	and IRIE indicate the amount of memory occupied by corresponding
	algorithms.
For the WC model, IRIE is much more efficient in terms of memory than PMIA
	for all the datasets.
The memory usages of PMIA are 4-20 times larger than the size of raw data
	file and also 2-7 times larger than that of IRIE.
Especially, for the LiveJournal dataset, PMIA requires about 10GB of memory
	spaces while IRIE requires only 3GB of memory which is close to the size
	of the raw text file.
We observe the similar patterns in memory usage for the TR model.
However, the amounts of memory occupied by PMIA are even larger than the WC
	model while the memory usages of IRIE are almost same with those for the
	WC model.
For the LiveJournal, PMIA requires about 16GB of memory which is an
	infeasibly large amount of memory.

%
%
%

%
\begin{figure}
\centering
\numberwithin{figure}{section}
\subfigure[Arxiv-WC]{\includegraphics[width = 0.35 \textwidth]{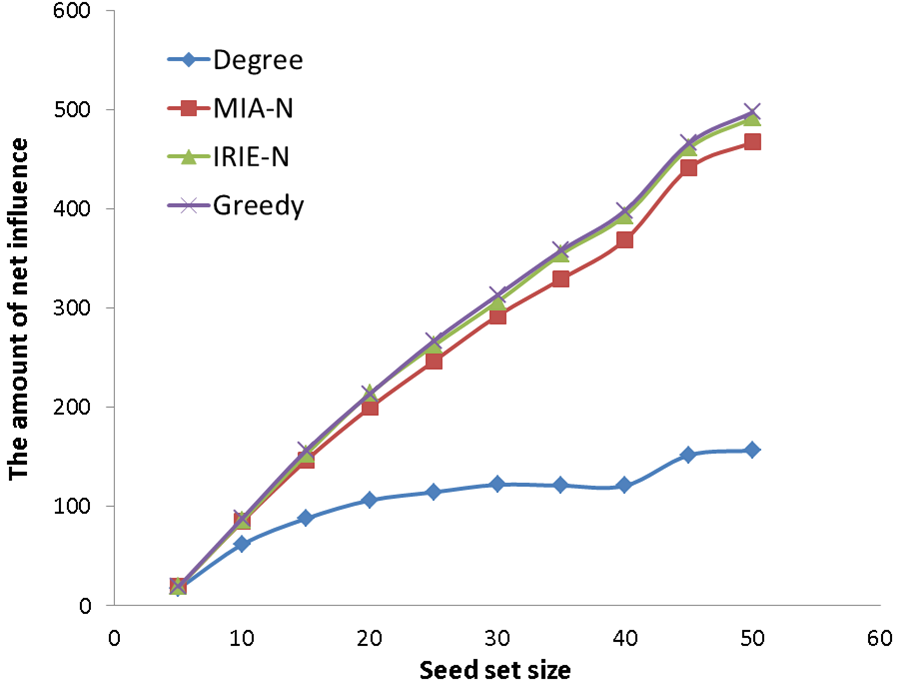}}
\subfigure[Arxiv-TR]{\includegraphics[width = 0.35 \textwidth]{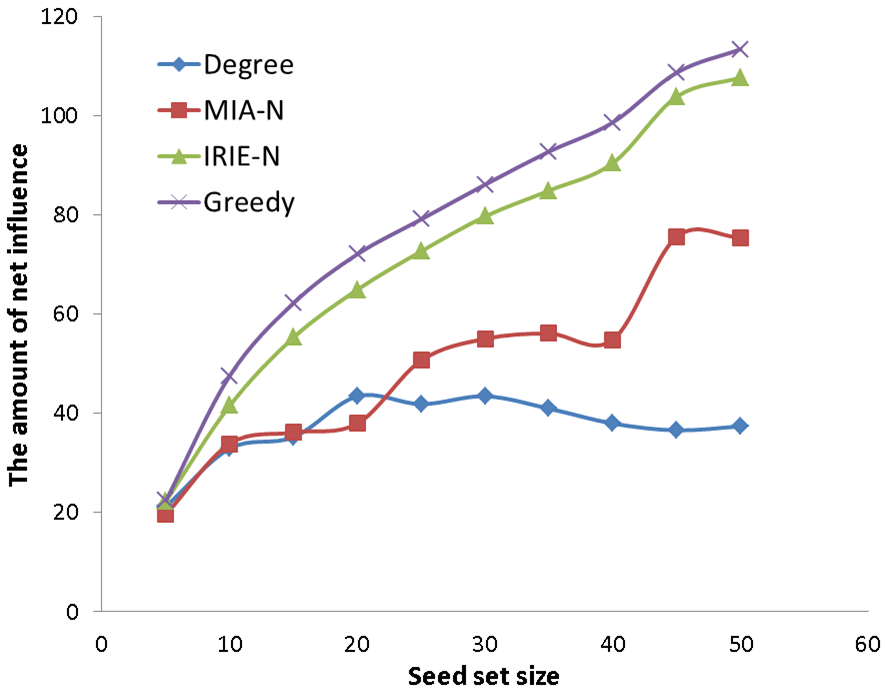}}
\\
\subfigure[Wiki-WC]{\includegraphics[width = 0.35 \textwidth]{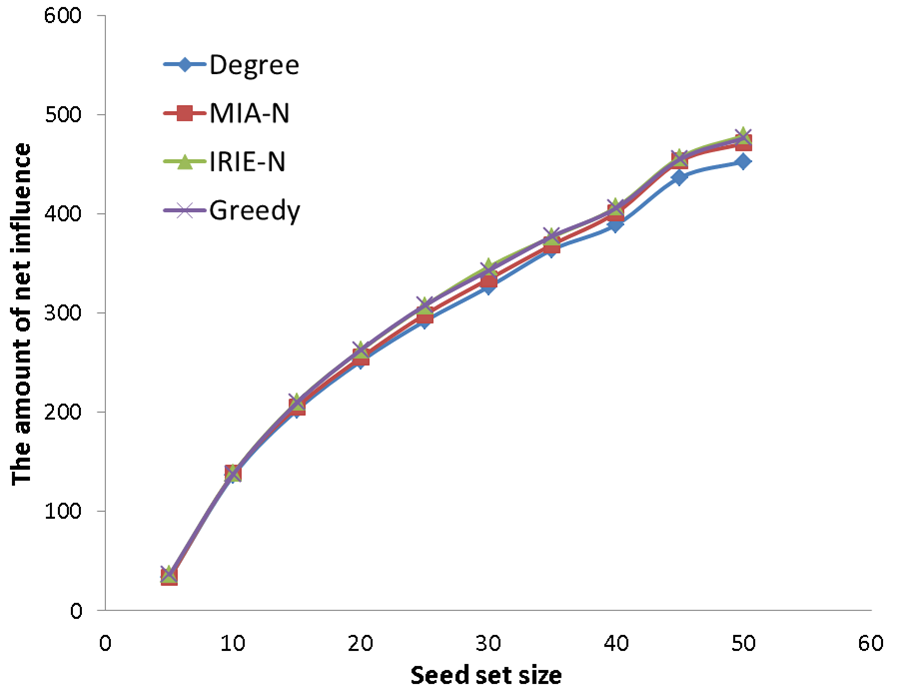}}
\subfigure[Wiki-TR]{\includegraphics[width = 0.35 \textwidth]{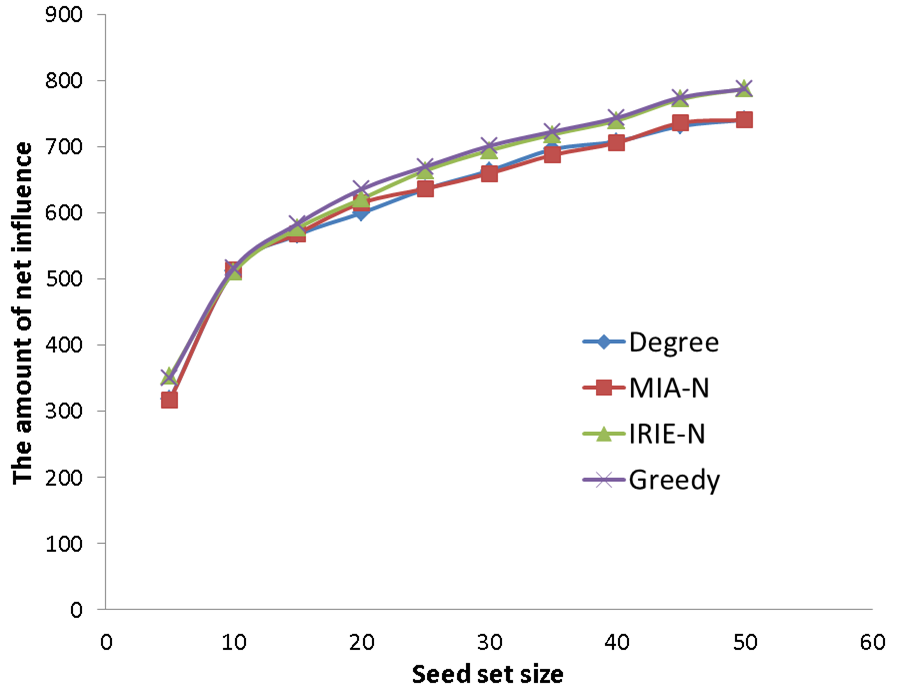}}
\caption{Influence spreads for IC-N model with q = 0.9, $\lambda$ = 0}
\label{fig:IC-N}
\end{figure}
\begin{figure}
\centering
\numberwithin{figure}{section}
\includegraphics[width = 0.45 \textwidth]{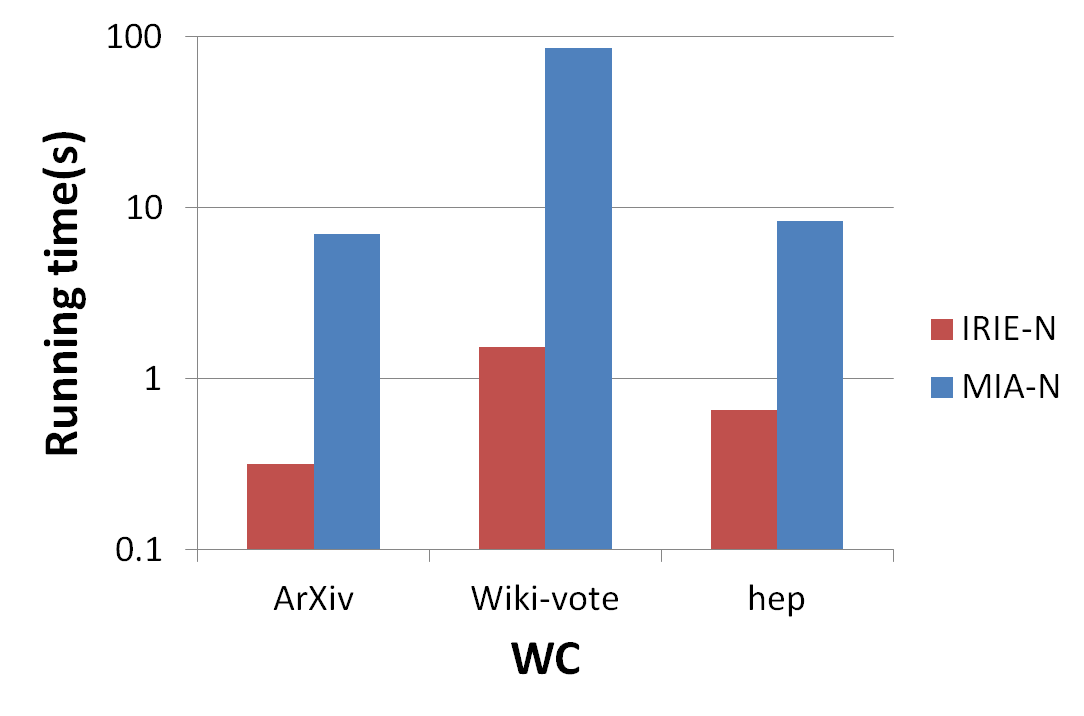}
\caption{Running time of algorithms under IC-N model where q = 0.9, $\lambda$ = 0}
\label{fig:running2}
\end{figure}

\subsection{Experiments on IC-N Model}
In this subsection, we show experimental results for the IC-N model.
Figure \ref{fig:IC-N} (a)-(d) show the influence spread of the algorithms.
When $\lambda = 0$, Greedy-N and IRIE-N shows the best performances, while MIA-N shows slightly less performance than IRIE-N. Note that for Arxiv-TR, IRIE-N shows more stable influence spread than MIA-N.
For the running time described in Figure \ref{fig:running2}, \IRIE-N is about 5-50 times faster than MIA-N. Hence we conclude that IRIE-N is much faster than other algorithms while achieving best influence spread.

\section{Conclusion}
In this paper, we propose a new scalable and robust algorithm IRIE for influence
	maximization under the independent cascade (IC) model and its extension
	IC-N model.
The IRIE algorithm incorporates fast iterative ranking algorithm (IR) with a
	fast influence estimation (IE) method to achieve scalability and robustness while
	maintaining good influence coverage.
Comparing with other state-of-the-art influence maximization algorithms, 
the advantage of IRIE is that it avoids the storage and
	computation of local data structures, which results in significant
	savings in both memory usage and running time.
Our extensive simulations results on synthetic and real-world networks
	demonstrate that IRIE is the best in influence coverage among
	all tested heuristics including PMIA, SAEDV, PageRank, degree heuristic, etc.,
	and it achieves up to two orders of magnitude speed-up with only
	a small fraction of memory usage, especially on
	relatively dense social networks (with average degree greater than $10$),
	comparing with other state-of-the-art heuristics.

An additional advantage of IRIE is that its simple iterative computation
	can be readily ported to a parallel graph computation platform
	(e.g. Google's Pregel~\cite{Pregel}) to further scale up influence
	maximization, while other heuristics such as PMIA involves more
	complicated data structures and is relative harder for parallel
	implementation.
A future direction is thus validating and improving the IRIE algorithm on a
	parallel graph computation platform.
Another future direction is to apply the IRIE framework
	to other influence diffusion models, such as the linear threshold model. 
\vspace{0.5cm}

%
\bibliographystyle{abbrv}
\bibliography{ref}  

\begin{thebibliography}{10}

\bibitem{BP98}
S.~Brin and L.~Page.
\newblock The anatomy of a large-scale hypertextual web search engine.
\newblock {\em Computer Networks}, 1998.

\bibitem{Chen:IC-N}
W.~Chen, A.~Collins, R.~Cummings, T.~Ke, Z.~Liu, D.~Rincon, X.~Sun, Y.~Wang,
  W.~Wei, and W.~Yuan.
\newblock Influence maximization in social networks when negative opinions may
  emerge and propagate.
\newblock In {\em SDM}, 2011.

\bibitem{Chen:PMIA}
W.~Chen, C.~Wang, and Y.~Wang.
\newblock Scalable influence maximization for prevalent viral marketing in
  large-scale social networks.
\newblock In {\em KDD}, 2010.

\bibitem{Chen:EIM}
W.~Chen, Y.~Wang, , and S.~Yang.
\newblock Efficient influence maximization in social networks.
\newblock In {\em KDD}, 2009.

\bibitem{CYZ10}
W.~Chen, Y.~Yuan, and L.~Zhang.
\newblock Scalable influence maximization in social networks under the linear
  threshold model.
\newblock In {\em ICDM}, 2010.

\bibitem{Domingos:MNVC}
P.~Domingos and M.~Richardson.
\newblock Mining the network value of customers.
\newblock In {\em KDD}, 2001.

\bibitem{data}
A.~Goyal, F.~Bonchi, and L.~V.~S. Lakshmanan.
\newblock A data-based approach to social influence maximization.
\newblock In {\em PVLDB}, 2011.

\bibitem{Goyal:CELF++}
A.~Goyal, W.~Lu, and L.~V.~S. Lakshmanan.
\newblock Celf++: optimizing the greedy algorithm for influence maximization in
  social networks.
\newblock In {\em WWW(Companion Volume)}, 2011.

\bibitem{JSCWSX11}
Q.~Jiang, G.~Song, G.~Cong, Y.~Wang, W.~Si, and K.~Xie.
\newblock Simulated annealing based influence maximization in social networks.
\newblock In {\em AAAI}, 2011.

\bibitem{Kempe:MSI}
D.~Kempe, J.~Kleinberg, and E.~Tardos.
\newblock Maximizing the spread of influence through a social network.
\newblock In {\em KDD}, 2003.

\bibitem{Kimura:TID}
M.~Kimura and K.~Saito.
\newblock Tractable models for information diffusion in social networks.
\newblock In {\em PKDD}, pages 259--271. LNAI 4213, 2006.

\bibitem{Leskovec:SNAP}
J.~Leskovec.
\newblock http://snap.stanford.edu/index.html.

\bibitem{Leskovec:CELF}
J.~Leskovec, A.~Krause, C.~Guestrin, C.~Faloutsos, J.~VanBriesen, and N.~S.
  Glance.
\newblock Cost-effective outbreak detection in networks.
\newblock In {\em KDD}, 2007.

\bibitem{Pregel}
G.~Malewicz, M.~H. Austern, A.~J.~C. Bik, J.~C. Dehnert, I.~Horn, N.~Leiser,
  and G.~Czajkowski.
\newblock Pregel: a system for large-scale graph processing.
\newblock In {\em SIGMOD}, 2010.

\bibitem{SuriN10}
R.~Narayanam and Y.~Narahari.
\newblock A shapley value based approach to discover influential nodes in
  social networks.
\newblock {\em IEEE Transactions on Automation Science and Engineering}, 2010.

\bibitem{Rozin:NB}
P.~Rozin and E.~B. Royzman.
\newblock Negativity bias, negativity dominance, and contagion.
\newblock {\em Personality and Social Psychology Review}, 2001.

\bibitem{Wang:CGA}
Y.~Wang, G.~Cong, G.~Song, , and K.~Xie.
\newblock Community-based greedy algorithm for mining top-k influential nodes
  in mobile social networks.
\newblock In {\em KDD}, 2010.

\end{thebibliography}

\newpage
\newpage
\section*{Appendix}
\appendix
\section{Proof of Theorem 2}
\begin{proof} 
First, note that Algorithm \ref{BP} computes the unique solution of
	(\ref{e_gm}) and (\ref{e_m}).
Let $m_t(u, v)$ be the expected number of activated nodes when $S = \{u\}$ and $u$
	activates other nodes within distance $t$ from $u$ using all out-going edges 
	of $u$ except for $(u, v)$.
Let $\tilde{m}_t(u, v)$ be the computed values from Algorithm \ref{BP}. Then we 
	will prove that $\tilde{m}_t(u, v) = m_t(u, v)$ for all $t = 0, 1, 2, \ldots$
	by a mathematical induction.
When $t = 0$, $\tilde{m}_0(u, v) = m_0(u, v) = 0$ for each edge $(v, u) \in E$.

Suppose that the statement is true for all $t < T$. Let $t = T$, and fix $u \in V$.
	Let $T_u$ be the tree graph $G$ whose root is $u$, and for each $w \in N^{out}
	(u)$, let $T_{uw}$ be the subtree of $T_u$ whose root is $w$.
Note that $m_{t-1}(w, u)$ is the expected influence of $\{w\}$ to the nodes in
	$T_{uw}$ within distance $t-1$ from $w$.
	
Since $T_u$ is a tree graph, by the linearity of expectation and the definition of
	$m_t(u, v)$, we have for any $(v, u) \in E$, 
\begin{equation}
\label{e_am}
m(u, v) = 1 + \left(\sum_{w \in N^{out}(u), w \ne v} P_{uw} \cdot m(w, u)\right).
\end{equation}
From the line \ref{alg:getm} of Algorithm \ref{BP}, for any $(v, u) \in E$,
\begin{equation}
\label{e_atm}
\tilde{m}_t(u, v) = 1 + \left(\sum_{w \in N^{out}(u), w \ne v} P_{uw} \cdot \tilde{m}_{t-1}(w, u)\right).
\end{equation}
From the induction hypothesis, $\tilde{m}_t(w, u) = m_{t-1}(w, u)$. Hence, from 
	(\ref{e_am}) and (\ref{e_atm}), we have that for any $(v, u) \in E$, 
	$m_t(u, v) = \tilde{m}_t(u, v)$.
Therefore we have shown the induction. 
Note that $m(u, v) = m_{|V|-1}(u, v)$ since the longest shortest path of $G$ has length
	at most $|V| - 1$. Hence $\{m_t(u, v)\}_t$ converges before $t \leq |V|$, and the
	same holds for $\{\tilde{m}_t(u, v)\}_t$.
	
Since $\{\tilde{m}_t(u, v)\}$ are the converged values $\{\tilde{m}_t(u, v)\}_t$ by
	the line \ref{alg:getm} of Algorithm \ref{BP}, we have that $\tilde{m}(u, v) =
	\tilde{m}_{|V|}(u, v) = m_{|V|}(u, v) = m(u, v)$ for all $(v, u) \in E$.

Since $G$ is a tree, from the definition of $\sigma(u)$ and the linearity of 
	expectation, 
\begin{displaymath}
\sigma(u) = 1 + \sum_{v \in N^{out}(u)} P_{uv} \cdot m(v, u).
\end{displaymath}

Here from (\ref{e_gm}), $\tilde{\sigma}(u) = \sigma(u)$ for all $u \in V$.
\end{proof}

\end{document}